\documentclass[letterpaper,english,aps,prl,superscriptaddress,twocolumn,amsfonts, amssymb]{revtex4}
\usepackage[T1]{fontenc}
\pdfoutput=1
\usepackage{textcomp}
\usepackage{mathrsfs}
\usepackage{amsmath}
\usepackage{amssymb}
\usepackage{graphicx}
\usepackage{esint}

\makeatletter
\usepackage[english]{babel}

\usepackage[bookmarks=true,colorlinks,linkcolor=blue,urlcolor=blue,citecolor=blue]{hyperref}
\usepackage{xr}
\usepackage{epsfig,graphicx,psfrag,amsmath,amssymb,float}
\usepackage[caption=false]{subfig}

\@ifundefined{showcaptionsetup}{}{%
 \PassOptionsToPackage{caption=false}{subfig}}
\usepackage{subfig}
\makeatother

\usepackage{babel}

\begin{document}
\title{Dynamics of a two-dimensional quantum spin-orbital liquid: \\ spectroscopic signatures
of fermionic magnons}
\author{Willian M. H. Natori}
\affiliation{Blackett Laboratory, Imperial College London, London SW7 2AZ, United
Kingdom}
\author{Johannes Knolle}
\affiliation{Department of Physics, TQM, Technische Universit\"{a}t \"{M}unchen,
85748 Garching, Germany}
\affiliation{Munich Center for Quantum Science and Technology (MCQST), 80799 Munich,
Germany}
\affiliation{Blackett Laboratory, Imperial College London, London SW7 2AZ, United
Kingdom}
\begin{abstract}
We provide an exact study of dynamical correlations
for the quantum spin-orbital liquid phases of an
 SU(2)-symmetric Kitaev honeycomb lattice model. We show that the spin
dynamics in this Kugel-Khomskii type model is exactly the density-density correlation
function of $S=1$ fermionic magnons, which could be probed in resonant
inelastic x-ray scattering (RIXS) experiments. We predict the characteristic signatures of spin-orbital fractionalization in inelastic scattering experiments and compare them to the ones of the spin-anisotropic Kitaev honeycomb spin liquid. In particular, the RIXS response shows a characteristic momentum dependence directly related to the dispersion of fermionic excitations. The Neutron scattering cross section displays a mixed response of fermionic magnons as well as spin-orbital excitations. The latter has a vison gap and a bandwidth of broad excitations, which is three times larger than the one of the spin-1/2 Kitaev model.  
\end{abstract}
\maketitle

Phases of matter which remain disordered down to lowest temperatures because of quantum fluctuations have fascinated the condensed matter community for a long time. One reason for this enduring interest is that they can host long-range entangled ground states displaying topological order~\cite{Wen1991}. Quantum spin liquids (QSLs)  ~\citep{Knolle2019,Savary2016} are prominent examples of such phases that have not been conclusively identified in experiment, in spite of the availability of many candidate materials. In addition to the absence of local order, the main reason for this long ongoing search for a QSL is  the unusual nature of its excitations which carry only fractions of the usual quantum numbers probed experimentally. For example, an $S=1$ spin flip excitation, diagnosed via the dynamical structure factor (DSF) in inelastic neutron scattering (INS), decays into multiple excitations, e.g. spinons and visons~\cite{punk2014topological} or Majorana fermions and fluxes~\cite{Knolle2014}, leading only to a broad featureless continuum response. An additional obstacle in this ongoing search is the fact that quantum liquids are inherently strongly interacting which makes it difficult to obtain rigorous theoretical predictions that could be compared to experiments beyond one-dimensional model cases.  

An important conceptual  development was the advent of exactly soluble models with QSL phases. The most prominent is the Kitaev honeycomb lattice model~\citep{Kitaev2006,Hermanns2018,motome2020hunting}, which has permitted the calculation of exact results for dynamical correlations in the thermodynamic limit as probed in scattering experiments~\cite{Baskaran2007,Knolle2014,Knolle2014b}. The fractionalized excitations of the Kitaev spin liquid (KSL) are Majorana fermions in a plaquette flux background. The prediction that the Kitaev model could be relevant to specific heavy-ion Mott insulators (the\emph{ Kitaev materials}) \citep{Jackeli2009}
rapidly followed by their synthesis~\citep{Winter2017,Hermanns2018,Takagi2019} provided additional motivation to evaluate dynamical 
response functions of a variety of scattering experiments ~\citep{Knolle2014,Knolle2014b,Knolle2015,Smith2015,Perreault2015,Knolle2016,Smith2016,Perreault2016b,Halasz2016,Halasz2017,Nasu2014,Nasu2015,Nasu2017}.
Unfortunately, most Kitaev materials~\citep{Winter2017,Hermanns2018,Takagi2019} show residual long range magnetic order instead of a pristine KSL phase, an observation well-explained by more complete models beyond the pure Kitaev limit~\citep{Chaloupka2013,Rau2014,Winter2016}. Nevertheless, the main features of the INS response of the Kitaev candidate material $\alpha$-RuCl$_{3}$ is arguably captured by the DSF of the Kitaev
model~\citep{Banerjee2016,Banerjee2017,do2017majorana,Banerjee2018,Knolle2018A}. In spite of these recent developments, our understanding -- even of the basic phenomenology and experimental signatures -- of quantum liquids beyond the pure Kitaev model remains limited. 

Here, we provide exact results for the dynamical response of a quantum spin-orbital liquid (QSOL) as found in certain Kugel-Khomskii (KK) models~\citep{Yao2009,Nussinov2009,Wu2009,Chua2011,Yao2011}. 
We focus on systems with four degrees of freedom per site which are either
 equivalent to $j=3/2$ spin models or KK models with
doubly degenerate orbitals~\cite{Nussinov2015,Carvalho2018}. Thereby, we uncover qualitative differences to QSLs of the anisotropic $j=1/2$ Kitaev type. In particular, we show that in a QSOL a $S=1$ spin flip can excite only one type of excitation, e.g. two Majorana fermions without additional fluxes, leading to a much cleaner signature of fractionalization with a distinct momentum dependence absent in the KSL.

We compute the dynamical correlation functions of
the SU(2)-symmetric Kitaev model
\begin{equation}
H=-\sum_{\left\langle lm\right\rangle _{\gamma}}J_{\gamma}\left(T_{l}^{\gamma}\boldsymbol{\sigma}_{l}\right)\cdot\left(T_{m}^{\gamma}\boldsymbol{\sigma}_{m}\right),\label{eq:MainModel}
\end{equation}
which is a generalization of the spin anisotropic Kitaev model~\citep{Yao2011}. Here, $J_{\gamma}$ are bond-dependent exchange constants, $\mathbf{T}$
and $\boldsymbol{\sigma}$ are orbital and spin operators satisfying
$\left[T_{l}^{\alpha},T_{m}^{\beta}\right]=2i\delta_{lm}\epsilon^{\alpha\beta\gamma}T_{l}^{\gamma}$,
$\left[\sigma_{l}^{\alpha},\sigma_{m}^{\beta}\right]=2i\delta_{lm}\epsilon^{\alpha\beta\gamma}\sigma_{l}^{\gamma}$
and $\left[T_{l}^{\alpha},\sigma_{m}^{\beta}\right]=0$. The model is another rare example of an
 exactly soluble one using a Majorana fermion representation of the spin-orbital
operators~\citep{Wang2009,Yao2011}. It displays a QSOL ground state
with an emergent $Z_{2}$ gauge field and fermionic excitations of the Majorana type related to spin flips dubbed
\emph{fermionic magnons} \citep{Yao2011}. 

We find that the dynamical response of spin
operators is given by the the density-density correlation $I(\mathbf{q},\omega)$
of fermionic excitations, which can be probed with resonant inelastic
x-ray scattering (RIXS) if $H$ is regarded as a $j=3/2$ model \citep{Natori2017}.
The DSF of the model is a linear combination of the flux diagonal part $I(\mathbf{q},\omega)$ and
a correlation function among the operators $\sigma^{\alpha}T^{\beta}$ exciting both types of excitations.

\emph{The model }- The SU(2)-symmetric Kitaev model has a macroscopic
set of conserved plaquette operators $\hat{W}_{p}$ analogous to the
ones in the spin-1/2 Kitaev model \citep{Kitaev2006,Yao2011}. A key
difference is that each $\hat{W}_{p}$ affects only the orbital degrees
of freedom of Eq. (\ref{eq:MainModel}) and trivially commutes with
all spin operators. The ground state of $H$ is easily found in an
enlarged Hilbert space defined by a six-flavor Majorana representation
of $\boldsymbol{\sigma}$ and $\mathbf{T}$: $\sigma_{i}^{\alpha}=-\frac{i}{2}\epsilon^{\alpha\beta\gamma}\eta_{i}^{\beta}\eta_{i}^{\gamma}$
and $T_{i}^{\alpha}=-\frac{i}{2}\epsilon^{\alpha\beta\gamma}\theta_{i}^{\beta}\theta_{i}^{\gamma}$
\citep{Wang2009,Yao2011}. The physical states are eigenstates of
the projector $D_{i}=i\eta_{i}^{x}\eta_{i}^{y}\eta_{i}^{z}\theta_{i}^{x}\theta_{i}^{y}\theta_{i}^{z}$
with eigenvalue $+1$. This constraint also entails that $\sigma_{i}^{\alpha}T_{i}^{\beta}=-i\eta_{i}^{\alpha}\theta_{i}^{\beta}$
and allows to represent Eq. (\ref{eq:MainModel}) like \citep{Yao2011}

\begin{equation}
\mathcal{H}=\sum_{\left\langle ij\right\rangle _{\gamma}}\sum_{\alpha}J_{\left\langle ij\right\rangle _{\gamma}}\hat{u}_{\left\langle ij\right\rangle _{\gamma}}i\eta_{i}^{\alpha}\eta_{j}^{\alpha},\label{eq:parton-SU2-Kitaev}
\end{equation}
where $\hat{u}_{\left\langle ij\right\rangle _{\gamma}}=i\theta_{i}^{\gamma}\theta_{j}^{\gamma}$
is a $\mathbb{Z}_{2}$ gauge operator defined along the bond $\left\langle ij\right\rangle _{\gamma}$
with $i$ on the even sublattice.

Note, Eq. (\ref{eq:parton-SU2-Kitaev}) generalizes the fermionic representation of the spin-1/2 Kitaev model~\citep{Kitaev2006}
by the presence of {\it three} Majorana flavors instead of one. Any
eigenstate $\left|\psi\right\rangle $ of $\mathcal{H}$ is then a
direct product $\left|\psi\right\rangle =\left|F_{\psi}\right\rangle \otimes\prod_{\alpha}\left|M_{\psi}^{\alpha}\right\rangle \equiv\left|F_{\psi}\right\rangle \otimes\left|M_{\psi}\right\rangle $,
where $\left|F_{\psi}\right\rangle $ is the flux sector and $\left|M_{\psi}^{\alpha}\right\rangle $
is a state for the $\eta^{\alpha}$ Majorana flavor of the ``matter''
sector. Lieb's theorem \citep{Lieb1994} asserts that the global ground
state is found in the flux sector $\left|F_{0}\right\rangle $ characterized
by $\hat{W}_{p}\left|F_{0}\right\rangle =\left|F_{0}\right\rangle $
for all plaquettes. In the language of $\mathbb{Z}_{2}$ gauge operators,
$\left|F_{0}\right\rangle $ is obtained after fixing $u_{\left\langle ij\right\rangle _{\gamma}}=1$
for all gauge fields.
With a superposition of two Majoranas of the same flavor $\alpha$ but on different sublattices  within a unit cell, the Hamiltonian in a fixed gauge configuration can be written in terms of complex matter fermions. The translational symmetry of the ground state allows us then to introduce the Fourier transformation of these complex fermions $a_{\mathbf{q}}^{\alpha \dagger}$ corresponding to the matter excitations which diagonalize the Hamiltonian
\begin{equation}
\mathcal{H}_{0}=\sum_{\mathbf{q}}\sum_{\alpha}\left|\mu_{\mathbf{q}}\right|\left(2a_{\mathbf{q}}^{\alpha \dagger}a_{\mathbf{q}}^{\alpha}-1\right) \label{eq:diagonal_H0}
\end{equation}
where $\mu_{\mathbf{q}}=\sum_{\gamma}J_{\gamma}\exp\left(i\mathbf{q}\cdot\mathbf{n}_{\gamma}\right)$
with $\mathbf{n}_{x,y}=\left(\pm\frac{1}{2},\frac{\sqrt{3}}{2}\right)$
and $\mathbf{n}_{z}=\mathbf{0}$. 
Depending on the ratio of the exchange constants, the system describes gapped or gapless QSOLs as  the matter dispersions $\epsilon_{\mathbf{q}}=2\left|\mu_{\mathbf{q}}\right|$ is gapless for $\left|J_{z}\right|<\left|J_{x}\right|+\left|J_{y}\right|$ (and permutations) and gapped otherwise \citep{Kitaev2006}.

It is instructive to analyze the fractionalization processes implied
by Eq. (\ref{eq:parton-SU2-Kitaev}). We recall that the spin fractionalization
in the standard Kitaev model can be represented by $\sigma\sim em\epsilon$,
where $e$ and $m$ are visons corresponding to the insertion of $\pi$-fluxes
in two adjacent plaquettes and $\epsilon$ is the Majorana fermion
\citep{Kitaev2006,Baskaran2007,Savary2016}. The same kind of fractionalization
occurs here but now for the {\it spin-orbital} operators like $\sigma^{a}T^{b}\sim em\epsilon^{a}$.
In this case, the $e$ and $m$ particles only affect the
orbital sector and the three $\epsilon$ particles correspond to
the Majorana flavors $\eta^{\alpha}$ for spins. As a qualitatively new feature of the QSOL,
the spin $\sigma^{a}$ fractionalizes into two $\epsilon$ particles unrelated to the formation
of visons which we show in the following translates into qualitatively distinct features in the dynamical response.

\begin{figure*}
\begin{centering}
\includegraphics[width=0.6\columnwidth]{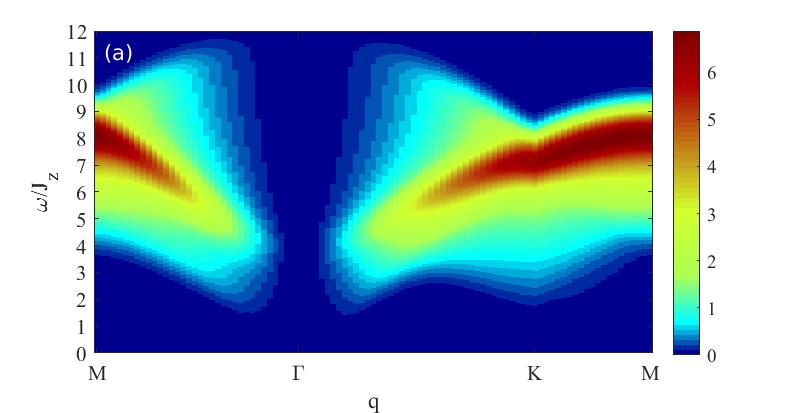}\includegraphics[width=0.6\columnwidth]{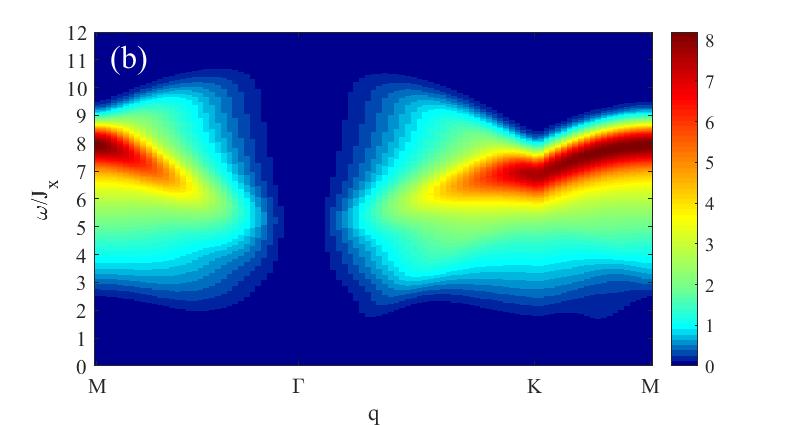}\includegraphics[width=0.6\columnwidth]{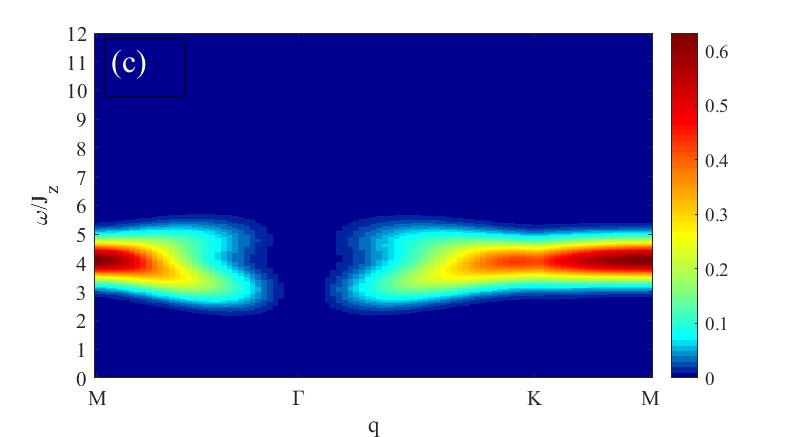}
\par\end{centering}
\begin{centering}
\includegraphics[width=0.6\columnwidth]{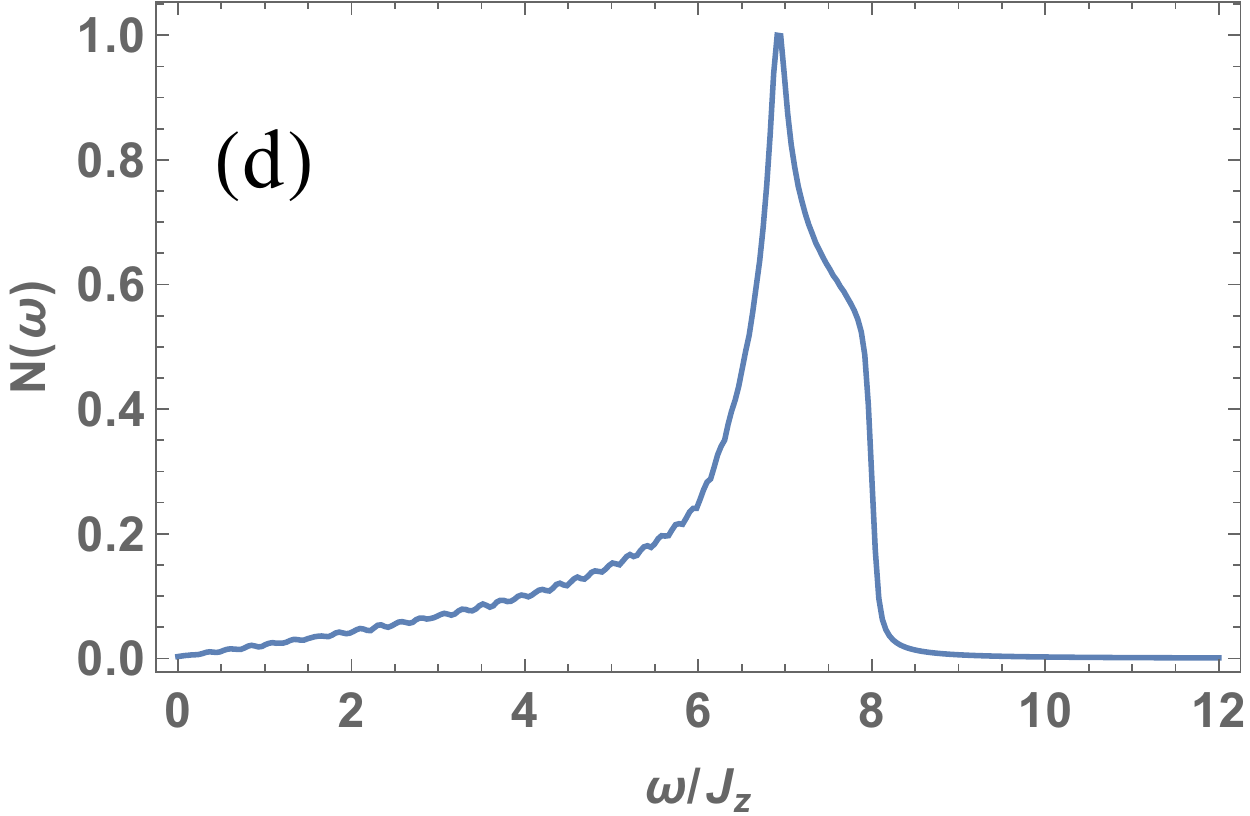}\includegraphics[width=0.6\columnwidth]{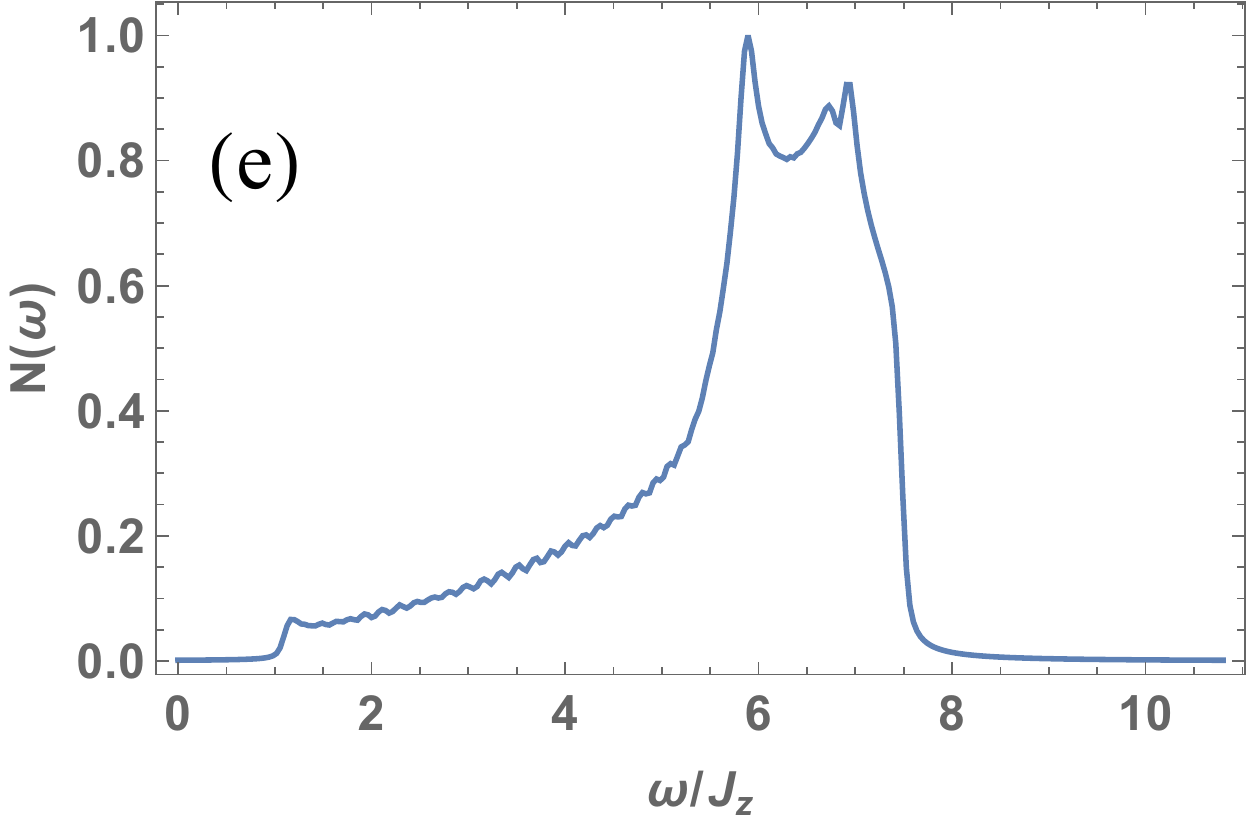}\includegraphics[width=0.6\columnwidth]{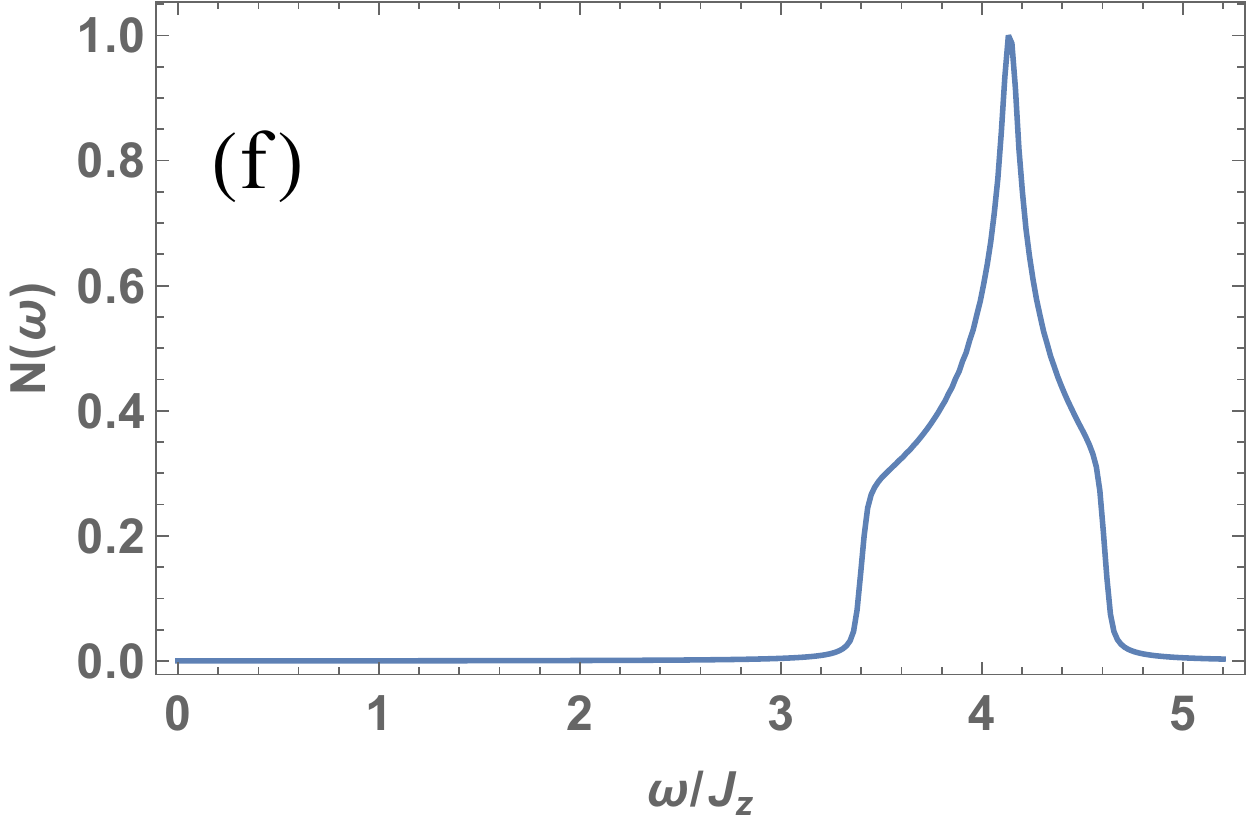}
\par\end{centering}
\caption{\label{fig:Sss_cases}(Color online) Dynamic structure factor $I\left(\mathbf{q},\omega\right)$ as measured in RIXS
and the normalized two-particle density of states $\rho(\omega)=\sum_{\mathbf{k}}\delta(\omega-\epsilon_{\mathbf{k}}-\epsilon_{\mathbf{k}+\mathbf{K}})$
for the cases (a+d) $J_{x}=J_{y}=J_{z}$, (b+e) $J_{x}=J_{y},J_{z}=0.7J_{x}$,
and (c+f) $J_{x}=J_{y}=0.15J_{z}$.}
\end{figure*}

\emph{Dynamical correlation functions} - We treat the SU(2)-symmetric
Kitaev model as a model of $j=3/2$ effective moments realized in
4/5$d^{1}$ Mott insulators, since this allows us to associate the
dynamical correlations to the expected responses of both RIXS and
INS~\citep{Natori2017}. Our goal is to compute the DSF given
by the Fourier transform of the correlation function $S_{lm}^{\alpha\beta}(t)=\left\langle \psi_{0}\left|j_{l}^{\alpha}(t)j_{m}^{\beta}(0)\right|\psi_{0}\right\rangle $,
of the angular momentum operators \citep{Natori2017}

\begin{align}
j_{l}^{\alpha} & \equiv-\frac{1}{2}\sigma_{l}^{\alpha}-\sigma_{l}^{\alpha}T_{l}^{\left(\alpha\right)},\label{eq:J_moments}
\end{align}
where $\alpha=x,y,z$, $T_{l}^{\left(z\right)}=T_{l}^{z}$ and $T_{l}^{\left(x,y\right)}=-\frac{1}{2}T_{l}^{z}\pm\frac{\sqrt{3}}{2}T_{l}^{x}.$
It turns out that the DSF is only a sum of two contributions because $\left\langle \sigma_{l}^{\alpha}(t)\sigma_{m}^{\beta}T_{m}^{\gamma}(0)\right\rangle =\left\langle \sigma_{l}^{\beta}T_{l}^{\gamma}(t)\sigma_{m}^{\alpha}(0)\right\rangle =0$,
since the action of $\sigma_{l}^{\alpha}T_{l}^{\left(\alpha\right)}$
on $\left|\psi_{0}\right\rangle $ involves the creation of a pair
of visons whereas $\sigma_{l}^{\alpha}$ is flux-conserving. We emphasize that the exact treatment discussed here considers the effects of these fluxes on the QSOL dynamics, which are not accounted for by standard mean-field treatments such as in Ref. \citep{Natori2017}. For comparison, the dynamics within a mean-field theory is provided in the Supplementary Material \citep{supp_ref}.

First, we discuss the correlation function $I_{ij}^{\alpha \beta}\left(t\right)=\left\langle \sigma_{i}^{\alpha}(t)\sigma_{j}^{\beta}(0)\right\rangle $.
The application of $\sigma_{j}^{\beta}$ on $\left|\psi_{0}\right\rangle $
preserves the gauge fluxes, thus allowing the evaluation of $I_{ij}^{\alpha \beta}\left(t\right)$
in terms of ground state correlations of Majorana fermions \citep{Carvalho2018}.
Additionally, since the Hamiltonian is diagonal in the Majorana flavor index we find $I_{ij}^{\alpha \beta}(t)\propto\delta_{\alpha \beta}$ 
and the SU(2) symmetry implies that $I_{ij}^{\alpha\alpha}\left(t\right)$ is isotropic for all $\alpha=x,y,z$. Hence, we only need to
evaluate a single (we omitted the $zz$-label)

\begin{align}
I_{lm}\left(t\right) & =-\sum_{\lambda}e^{i\left(E_{0}-E_{\lambda}\right)t}\left\langle M_{0}\left|\eta_{l}^{x}\eta_{l}^{y}\left|\lambda\right\rangle \left\langle \lambda\right|\eta_{m}^{x}\eta_{m}^{y}\right|M_{0}\right\rangle \label{eq:Izzt}
\end{align}
where the sum runs over all two-particle excitations of $\left|M_{0}^{x}\right\rangle $
and $\left|M_{0}^{y}\right\rangle $. A convenient representation
of Eq. (\ref{eq:Izzt}) is given in terms of $S=1$ fermionic magnons
defined by $c_{l}^{\dagger}=\frac{1}{2}\left(\eta_{l}^{x}+i\eta_{l}^{y}\right)$
\citep{Yao2011}. After performing the Fourier transform, the spin-spin
correlation reads
\begin{align}
I\left(\mathbf{q},\omega\right) & =\frac{8\pi}{N}\sum_{\lambda}\sum_{\mathbf{R}_{l},\mathbf{R}_{m}}e^{i\mathbf{q}\cdot\left(\mathbf{R}_{m}-\mathbf{R}_{l}\right)}\delta\left[\omega-\left(E_{\lambda}-E_{0}\right)\right]\nonumber \\
 & \qquad\quad\quad\times\left\langle M_{0}\left|n_{\mathbf{R}_{l}}^{c}\right|\lambda\right\rangle \left\langle \lambda\left|n_{\mathbf{R}_{m}}^{c}\right|M_{0}\right\rangle \nonumber \\
 & \equiv\frac{2\pi}{N}\sum_{\mathbf{k}}\delta\left(\omega-2\left(\mu_{\mathbf{k}}+\mu_{\mathbf{k}+\mathbf{q}}\right)\right)\nonumber \\
 & \qquad\quad\times\left|1-e^{2i\left(\theta_{\mathbf{k}+\mathbf{q}}-\theta_{\mathbf{k}}\right)}\right|^{2}\label{eq:Izz_omega}
\end{align}
where $n_{\mathbf{R}_{j}}^{c}$ is the total number of $c$ fermions
at the $\mathbf{R}_{l}$ unit cell and $e^{-2i\theta_{\mathbf{k}}}=\mu_{\mathbf{k}}/\left|\mu_{\mathbf{k}}\right|$.
The first equation shows that $I\left(\mathbf{q},\omega\right)$ is
readily interpreted as the density-density correlations of fermionic
magnons. In contrast to the KSLs, the real-space spin-spin correlations decay algebraically (exponentially)
for gapless (gapped) fermionic dispersion \citep{Yao2011}.
Remarkably, the splitting between spin and orbital degrees of
freedom allowed a simple, yet exact, expression for the dynamics of
a QSOL with longer range correlations.

\begin{figure*}
\begin{centering}
\includegraphics[width=0.7\columnwidth]{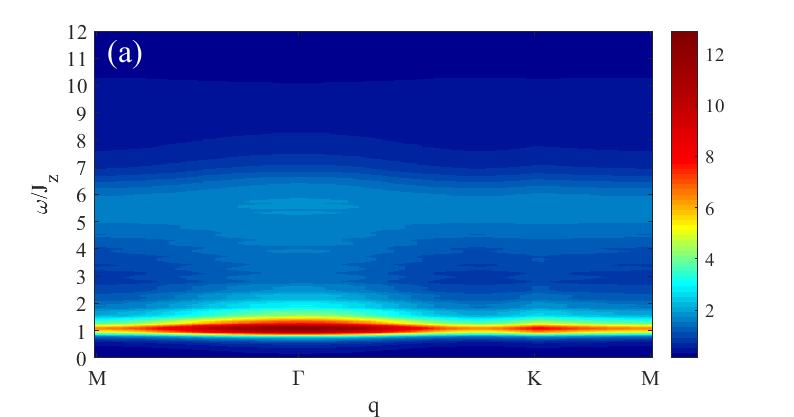}\includegraphics[width=0.7\columnwidth]{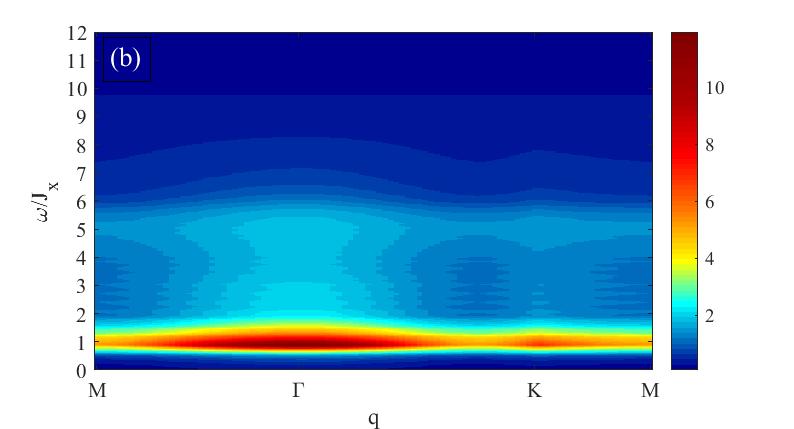}\includegraphics[width=0.7\columnwidth]{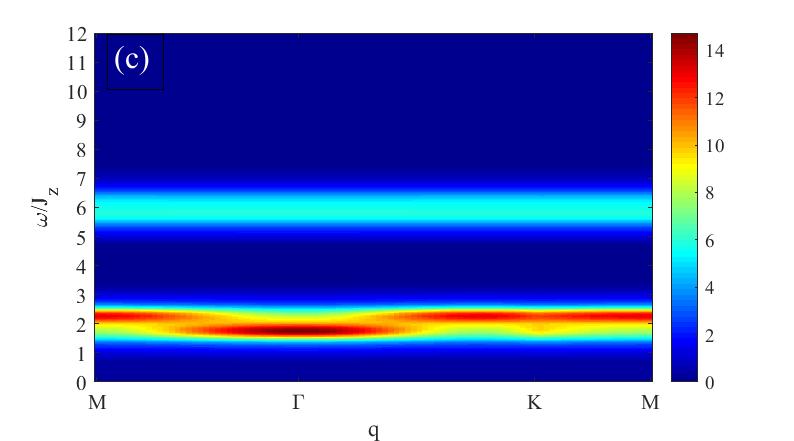}
\par\end{centering}
\begin{centering}
\includegraphics[width=0.7\columnwidth]{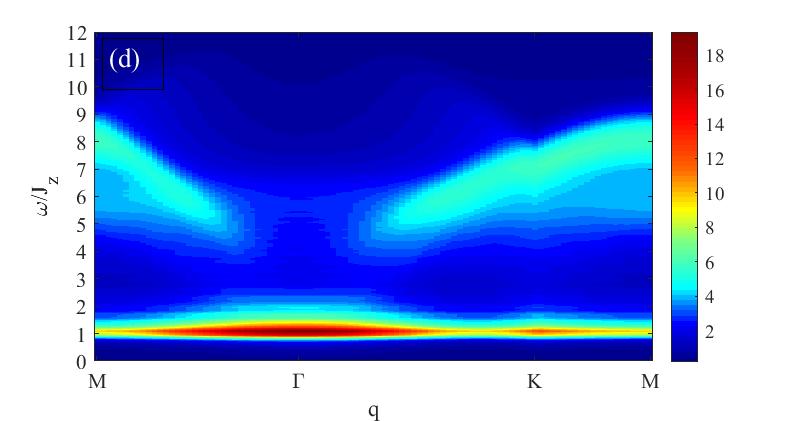}\includegraphics[width=0.7\columnwidth]{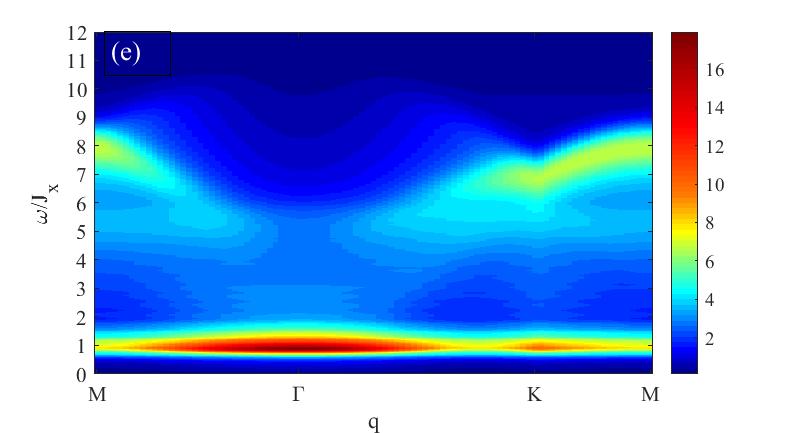}\includegraphics[width=0.7\columnwidth]{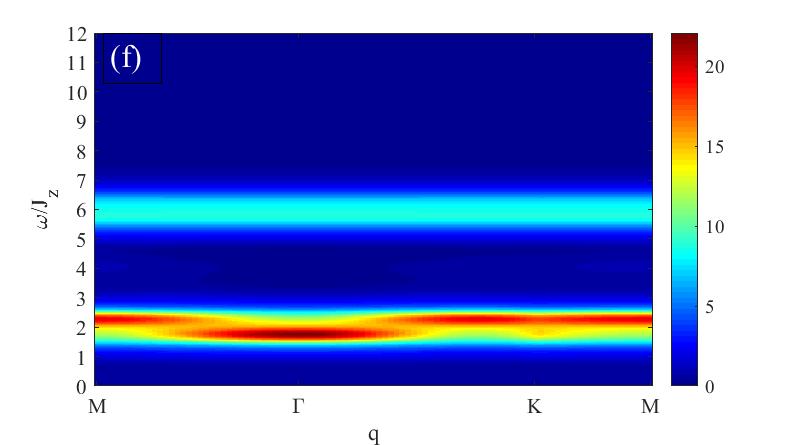}
\par\end{centering}
\caption{\label{fig:DSF_cases} The correlation function $\left(W_{z}+W_{x}\right)(\mathbf{q},\omega)$
(upper) and the DSF as measured in INS of the SU(2)-symmetric Kitaev model for the cases
(a+d) $J_{x}=J_{y}=J_{z}$, (b+e) $J_{x}=J_{y},J_{z}=0.7J_{x}$, and
(c+f) $J_{x}=J_{y}=0.15J_{z}$.}
\end{figure*}

Second, we show how the correlation functions $W_{lm}^{\alpha\beta,\delta\gamma}(t)=\left\langle \sigma_{i}^{\alpha}T_{i}^{\beta}(t)\sigma_{j}^{\delta}T_{j}^{\gamma}(0)\right\rangle $
are mapped into a quantum quench problem similar to the one discussed
for KSLs. Following the arguments of Ref. \citep{Baskaran2007}, $W_{lm}^{\alpha\beta,\delta\gamma}=\left\langle \sigma_{l}^{\alpha}T_{l}^{\beta}(t)\sigma_{m}^{\delta}T_{m}^{\gamma}(0)\right\rangle $
will be non-zero only if $l=m$ or if $l$ and $m$ are nearest neighbors.
Fixing $l$($m$) to the even (odd) sublattice, $W_{lm}^{\alpha\beta,\delta\gamma}(t)$
is obtained from 
\begin{align}
\left\langle \hat{\sigma}_{l}^{\alpha}T_{l}^{\beta}(t)\hat{\sigma}_{m}^{\delta}T_{m}^{\gamma}(0)\right\rangle  & =-i\left\langle M_{0}\left|\eta_{l}^{\alpha}e^{-i\left(\mathcal{H}_{0}+V_{\gamma}\right)t}\eta_{m}^{\delta}\right|M_{0}\right\rangle \nonumber \\
 & \quad\times e^{iE_{0}t}\delta^{\alpha\delta}\delta^{\beta\gamma}\delta_{\mathbf{r}(m),\mathbf{r}(l)+\mathbf{n}_{\gamma}},
\end{align}
in which $E_{0}$ is the ground state energy, $\mathbf{r}(l)$ is
the unit cell containing the site $l$, and $V_{\gamma}=-2J_{\gamma}i\sum_{\alpha}\left[\eta_{A,\mathbf{r}(l)}^{\alpha}\eta_{B,\mathbf{r}(l)+\mathbf{n}_{\gamma}}^{\alpha}\right]\equiv\sum_{\alpha}V_{\gamma}^{(\alpha)}$.
This expression differs from the quantum quenches obtained
in the spin-1/2 Kitaev model \citep{Knolle2014,Knolle2015,Smith2015,Knolle2016,Smith2016}
 by the number of flavors in $\mathcal{H}_{0}$, $V_{\gamma}$
and $\left|M_{0}\right\rangle $. Therefore, the non-zero matrix elements
have the form
\begin{align}
W_{lm,\gamma}^{\alpha\alpha}(t) & =\mathcal{W}_{lm,\gamma}^{\alpha\alpha}(t)\prod_{\delta\neq\alpha}L_{\gamma}^{\delta}(t)\delta_{\mathbf{r}(m),\mathbf{r}(l)+\mathbf{n}_{\gamma}}\label{eq:cor_sT_sT}
\end{align}
where 
\begin{align}
\mathcal{W}_{lm,\gamma}^{\alpha\alpha}(t) & =-ie^{iE_{0}t}\left\langle M_{0}^{\alpha}\left|\eta_{l}^{\alpha}e^{-i\left(\mathcal{H}_{0}^{(\alpha)}+V_{\gamma}^{(\alpha)}\right)t}\eta_{m}^{\alpha}\right|M_{0}^{\alpha}\right\rangle ,\nonumber \\
L_{\gamma}^{\delta}(t) & =\left\langle M_{0}^{\delta}\left|e^{-i\left(\mathcal{H}_{0}^{(\delta)}+V_{\gamma}^{(\delta)}\right)t}\right|M_{0}^{\delta}\right\rangle .\label{eq:quenches}
\end{align}
The matrix element in $\mathcal{W}_{lm,\gamma}^{\alpha\alpha}(t)$
is the same of the Kitaev model \citep{Knolle2014,Knolle2015,Knolle2016} but the multiple matter flavors result in a new time-dependent phase $L_{\gamma}^{\delta}(t)$ which can be calculated exactly via a Pfaffian formula from functional integrals \citep{Knolle2015}.
Finally, we exploit the SU(2) invariance of the model implying that $W_{lm,\gamma}^{\alpha\alpha}$
is flavor independent and focus on $\alpha=z$.

Overall, the DSF of the SU(2)-symmetric model is given by
\begin{equation}
S(\mathbf{q},\omega)=\frac{3}{4}I(\mathbf{q},\omega)+\frac{3}{2}\left(W_{z}+W_{x}\right)(\mathbf{q},\omega),\label{eq:DSF}
\end{equation}
where $W_{\gamma}(\mathbf{q},\omega)$ is the Fourier transform of
$W_{lm,\gamma}(t)$. Notice that correlations along the $y$-bonds
do not contribute to the DSF as predicted by the absence of the operators
$T_{l}^{y}$ in Eq. (\ref{eq:J_moments}). Physically, this reflects the
absence of coupling between the neutron spin and the $\sigma_{l}^{\alpha}T_{l}^{y}$
operators due to their evenness under time-reversal \citep{Natori2017,Yuan2018}.

\emph{Results} - In the following, we show the qualitatively different results of gapped and gapless QSOLs for three representative cases of Majorana dispersion: (i) gapless and isotropic ($J_{\alpha}=1$), (ii) gapless and anisotropic ($J_{z}<J_{x}=J_{y}$), and (iii) gapped ($J_{x}=J_{y}\ll J_{z}$). Let us first discuss the density-density correlation
of fermionic magnons $I\left(\mathbf{q},\omega\right)$ from dynamical spin correlations presented
in Eq. (\ref{eq:Izz_omega}). Note, for our choice of orbital representation it is directly measurable with
RIXS at the $L_{3}$-edge \citep{Natori2017}. The responses displayed
in Fig. \ref{fig:Sss_cases} strongly depend upon the value of the
transferred momenta $\mathbf{q}$ in contrast to the DSF of the spin-1/2
Kitaev model whose ultra short ranged spin correlations result in an almost dispersionless response \citep{Knolle2014,Knolle2015,Knolle2016}. 

An analysis
of $I\left(\mathbf{q}=\mathbf{K},\omega\right)$ shows that they closely
follow the density of states $\rho(\omega)$ of two-fermion excitations (lower panel), e.g. with intensity peaks related to the van Hove singularities of the fermionic bands.
In contrast to the gapped response of the spin-1/2
Kitaev model even for gapless fermions, one would expect a verifiable response of $I\left(\mathbf{q},\omega\right)$
for excitations below the vison gap in gapless QSOLs because of the different flux selection rules, especially when
$\mathbf{q}\approx\Gamma$. However, the form factor of $I\left(\mathbf{q},\omega\right)$
 vanishes at $\mathbf{q}=\Gamma$ which results in zero
 intensity at this point. This feature can be explained via the form factor at $\mathbf{q}=\Gamma$ which
is proportional to $\left|\left\langle \lambda\left|\sum_{i}\sigma_{i}^{z}\right|M_{0}\right\rangle \right|^{2}$.
Since $\left|M_{0}\right\rangle $ must be a many-body singlet of
$\boldsymbol{\sigma}$, $\sum_{i}\sigma_{i}^{z}\left|M_{0}\right\rangle =0$
and the response is zero \citep{Natori2017}.

It is interesting to note that the dynamical spin response $I\left(\mathbf{q},\omega\right)$ of the QSOL is similar to the RIXS response of the
spin-1/2 Kitaev model \citep{Halasz2016}. The form factor in both
cases is proportional to the term $\left|1-e^{2i\left(\theta_{\mathbf{k}+\mathbf{q}}-\theta_{\mathbf{k}}\right)}\right|^{2}$,
which is a direct consequence of the projective transformations of
fermions under inversion \citep{Halasz2016}. However, in the case of the spin-1/2 Kitaev
model, the form factor still arises from nearest neighbor correlations, which generates an additional factor $\left(\left|\mu_{\mathbf{k}}\right|-\left|\mu_{\mathbf{k}+\mathbf{q}}\right|\right)^{2}$.
Therefore, the response of the QSOL $I\left(\mathbf{q},\omega\right)$ has a stronger intensity at lower
energies and a more pronounced momentum dependence.

We now turn to the dynamical correlations of the spin-orbital operators displayed
in Fig. \ref{fig:DSF_cases}(a-c). The response $\left(W_{z}+W_{x}\right)(\mathbf{q},\omega)$
is qualitatively similar to the DSF of the spin-1/2 Kitaev model \citep{Knolle2014,Knolle2015,Knolle2016}.
There is a flux gap even in the gapless phase and only weak
dependence on the transferred momentum. However, there are
two important differences due to the additional Majorana flavors:
the flux gap is three times larger and the response extends to energies
beyond the Majorana fermion band width (shifted by the gap). These higher-energy excitations originate
from the extra phases $L_{\gamma}^{\delta}(t)$ in Eq. (\ref{eq:quenches}) and have a simple interpretation: the action of a spin-orbital operator excites one flavor of Majorana fermions and a pair of fluxes, the latter also shaking up the remaining two flavor sectors without fermion excitations resulting in the Loschmidt echo-like quench $L_{\gamma}^{\delta}(t)$.

Finally, the sum of the contributions, see Eq. (\ref{eq:DSF}), is the DSF as measurable in INS shown in Fig.\ref{fig:DSF_cases}(d-f). The DSF displays mixed characteristics
of the dynamics of fermionic magnons and the correlation of spin-orbital
operators. Our exact results provide a concrete example of how RIXS can complement studies
of INS to disentangle the different signatures of quantum number fractionalization related to the spin and orbital degrees of freedom in QSOLs. While RIXS measures the dispersion of fermionic excitations but not the flux gap, INS captures both features but is unable to distinguish them by itself.

\emph{Experimental connections} -
Ref. \citep{Yao2011} proposed that a decorated honeycomb
lattice can give rise to the SU(2)-symmetric Kitaev model but more
promising seems to be the connection with spin-orbital systems. The bond-frustrated exchanges of Eq. (\ref{eq:MainModel})
resembles those appearing in KK models \citep{Kugel1982,Nussinov2015} associated with Mott insulators that retain
$e_{g}$ degeneracy \citep{Kugel1982,Imada1998,Tokura2000,Khaliullin2005,Nussinov2015}.
The synthesis of $4/5d^{1}$ Mott insulators with $j=3/2$ magnetic
moments \citep{Chen2010,Natori2016,Romhanyi2017,Natori2017,Yamada2018,Natori2018}
or graphene-based superlattices \citep{Yuan2018,Venderbos2018,Natori2019b} has  
increased the list of Kugel-Khomskii materials. Finally, new routes to materials that implement the Kitaev model with 
higher spins have been proposed recently \citep{XuNPJ2018,Stavropoulos2019,Xu2020} and the methods developed here might be
useful to uncover their dynamics.

Interestingly, Eq. (\ref{eq:MainModel}) is expected to emerge in highly anisotropic materials, e.g. coupled chains, because of the inherent difference between spin and orbital operators. 
While the spin transforms as  $\Theta\boldsymbol{\sigma}\Theta^{-1}=-\boldsymbol{\sigma}$ under time-reversal $\Theta$ the orbital operators
$T^{x}$ and $T^{z}$ are time-reversal invariant and $\Theta T^{y}\Theta^{-1}=-T^{y}$
\citep{Natori2017,Yuan2018}. This symmetry property implies that
$\sigma^{\alpha}T^{x,z}$ must be a linear combination of dipoles
and octupoles of an effective $j=3/2$ angular moment while $\sigma^{\alpha}T^{y}$
are equivalent to quadrupoles of $\mathbf{j}$ \citep{Natori2017}.
The interactions along one of the bond directions is then of a different
nature in solid-state implementations of Eq. (\ref{eq:MainModel}). 

In general, the key ingredient of the model studied here is 
the SU(2) symmetry of spins which is common among several
KK models with possible QSOL ground states \citep{Chen2010,Natori2016,Romhanyi2017,Natori2017,Yamada2018,Natori2018,Venderbos2018,Natori2019b}.
Perturbations induced by Hund's coupling break this symmetry 
and will change the responses in realistic settings. At this point, we recall that the 
most prominent effect of similar perturbations to the isotropic spin-1/2 Kitaev model was to add a nonzero spectral weight in the neighborhood of the $\Gamma$ and $K$ points  \citep{Song2016}. A similar result is expected for the SU(2) extension of the model, but in this case the spectroscopic response at these points is finite already in the unperturbed limit. Overall, it would be desirable to study the quantitative effects of integrability breaking perturbations for the dynamics of QSOLs, for example by generalizing the augmented parton mean field theory developed for the spin 1/2 Kitaev model Ref. \citep{Knolle2018A}.   

\emph{Conclusion} - 
We provide the first exact results of dynamical correlations of a QSOL also giving an example for 
algebraically decaying spin liquids. Our computation of the dynamical spin- and orbital-correlations of an SU(2)-symmetric extension of the Kitaev model shows how spin-orbital fractionalization is manifest in scattering experiments like INS and RIXS. For example, it would be desirable to look for signatures of $S=1$ fermionic magnons with a distinct energy and momentum dependence in Kugel-Khomskii materials with SU(2) symmetry.

In the future it would be desirable to extend the as of yet short list of rigorous results for the dynamics (and finite temperature properties \citep{Nasu2014,Nasu2015,Nasu2017}) of quantum liquids to other exactly soluble systems, e.g. SU(2)-symmetric
Kitaev models on other tricoordinated lattices \citep{Smith2015,Smith2016,Halasz2017}, models with a spinon Fermi sea \citep{Yao2011,Zhang2019}, or those on four-coordinated lattices with half-integer spin per unit cell \citep{Yao2009,Nussinov2009,Wu2009,Chua2011,Whitsitt2012}
whose dynamical correlations are also mapped to quantum quench problems.

\emph{Acknowledgments} We acknowledge support from the Royal Society via a Newton International Fellowship through project NIF$\setminus$R1$\setminus$181696.

\bibliographystyle{apsrev4-1}
\bibliography{SU2Dyn2}

\newpage \leavevmode \newpage
\appendix

\onecolumngrid

\section{\large{Supplementary Material}}
\begin{center}
\textbf{Dynamics of a two-dimensional quantum spin-orbital liquid: \\ spectroscopic signatures
of fermionic magnons}\\ 
\vspace{10pt}
Willian Natori$^{1}$, Johannes Knolle$^{2,3,1}$ \\ \vspace{6pt}

$^1$\textit{\small{Blackett Laboratory, Imperial College London, London SW7 2AZ, United Kingdom}}
$^2$\textit{\small{Department of Physics and Institute for Advanced Study, Technical University of Munich, 85748 Garching, Germany}} \\
$^3$\textit{\small{Munich Center for Quantum Science and Technology (MCQST), Schellingstr. 4, D-80799 M{\"u}nchen, Germany}} \\
\end{center}

\maketitle

In this Supplementary Material, we develop a standard parton mean-field theory for
the SU(2)-symmetric Kitaev model

\begin{equation}
H=-\sum_{\gamma}\sum_{\left\langle ij\right\rangle _{\gamma}}J_{\gamma}T_{i}^{\gamma}T_{j}^{\gamma}\boldsymbol{\sigma}_{i}\cdot\boldsymbol{\sigma}_{j}\label{eq:SU(2)_model_supp}
\end{equation}
that recovers the exact results for the dispersion of the fermionic
magnons. We also uncover the dynamics of the model within this approximation
following the same methodology of Ref. \citep{Natori2017}. The main purpose is to highlight the differences with the exact solution presented in the main text. 

\section{Mean-field theory}

We first rewrite Eq. (\ref{eq:SU(2)_model_supp}) in terms of the
Majorana fermions introduced in the main text as follows

\begin{align}
H & =\sum_{\gamma}\sum_{\left\langle ij\right\rangle _{\gamma}}J_{\gamma}\left(i\theta_{i}^{\gamma}\theta_{j}^{\gamma}\right)\left(i\boldsymbol{\eta}_{i}\cdot\boldsymbol{\eta}_{j}\right).
\end{align}
From the knowledge of the exact solution, it is natural to define
the following order parameters
\begin{equation}
u_{ij}^{\gamma}=\left\langle i\theta_{i}^{\gamma}\theta_{j}^{\gamma}\right\rangle ,t_{ij}=\left\langle i\boldsymbol{\eta}_{i}\cdot\boldsymbol{\eta}_{j}\right\rangle .
\end{equation}
The order parameters display the same modulus throughout the whole
lattice but must obey the relations $u_{ji}^{\gamma}=-u_{ij}^{\gamma}$
and $t_{ji}=-t_{ij}$. Let us then fix our attention to the case in
which $i$ is on the even sublattice ($A$) and $j$ is on the odd
($B$) one. The mean-field Hamiltonian is then given by
\begin{align}
H_{\text{MF}} & =\sum_{\gamma}\sum_{\mathbf{r}}J_{\gamma}u_{i_{A}j_{B}}^{\gamma}\left(i\boldsymbol{\eta}_{\mathbf{r},A}\cdot\boldsymbol{\eta}_{\mathbf{r}+\mathbf{n}_{\gamma},B}\right)+\sum_{\gamma}\sum_{\mathbf{r}}J_{\gamma}t_{i_{A}j_{B}}\left(i\theta_{\mathbf{r},A}^{\gamma}\theta_{\mathbf{r}+\mathbf{n}_{\gamma},B}^{\gamma}\right)-\sum_{\gamma}\sum_{\mathbf{r}}J_{\gamma}u_{i_{A}j_{B}}^{\gamma}t_{i_{A}j_{B}},\label{eq:HMF}
\end{align}
where the nearest-neighbor vectors $\mathbf{n}_{\gamma}$ are explicitly
given by

\begin{align}
\mathbf{n}_{x} & =\frac{1}{2}\hat{\mathbf{x}}+\frac{\sqrt{3}}{2}\hat{\mathbf{y}},\nonumber \\
\mathbf{n}_{y} & =-\frac{1}{2}\hat{\mathbf{x}}+\frac{\sqrt{3}}{2}\hat{\mathbf{y}},\nonumber \\
\mathbf{n}_{z} & =\mathbf{0}.
\end{align}
Every Majorana flavor $\zeta_{i}^{\alpha}\in\left\{ \theta_{i}^{\gamma},\eta_{i}^{\gamma}\right\} $
satisfy the anticommutation relation $\left\{ \zeta_{i}^{\alpha},\zeta_{j}^{\beta}\right\} =2\delta^{\alpha,\beta}\delta_{i,j}$.
Such Majorana operators can be related to operators defined in momentum
space by the following Fourier transform
\begin{align}
\zeta_{\mathbf{r},X} & =\sqrt{\frac{2}{N}}\sum_{\mathbf{q}\in\text{BZ}}e^{-i\mathbf{q}\cdot\mathbf{r}}\zeta_{\mathbf{q},X},
\end{align}
in which $\mathbf{r}$ labels the unit cells, $X=A,B$ labels the
sublattices, $N$ is the total number of unit cells and the sum runs
over the first Brillouin zone of the honeycomb lattice. The normalization
factor was chosen in such a way that the original anticommutation
relations are consistent with $\left\{ \zeta_{\mathbf{q},X},\zeta_{\mathbf{q}^{\prime},Y}\right\} =\delta_{\mathbf{q}^{\prime},-\mathbf{q}}\delta_{X,Y}$.
This algebra allows us to treat $\zeta_{\mathbf{q},X}$ as a canonical
fermion if we (i) constrain $\mathbf{q}$ to one half of the Brillouin
zone and (ii) assign $\zeta_{\mathbf{q},X}^{\dagger}=\zeta_{-\mathbf{q},X}$.
Under these constraints, the general hopping Hamiltonian is given
by

\begin{equation}
\sum_{\left\langle ij\right\rangle _{\gamma}}i\zeta_{i}\zeta_{j}=2i\sum_{\mathbf{q}\in\frac{1}{2}\text{BZ}}\left(e^{-i\mathbf{q}\cdot\mathbf{n}_{\gamma}}\zeta_{\mathbf{q},A}^{\dagger}\zeta_{\mathbf{q},B}-e^{i\mathbf{q}\cdot\mathbf{n}_{\gamma}}\zeta_{\mathbf{q},B}^{\dagger}\zeta_{\mathbf{q},A}\right),\label{eq:general_hopping}
\end{equation}
After replacing Eq. (\ref{eq:general_hopping}) on Eq. (\ref{eq:HMF})
and defining the spinor $\left(\zeta_{\mathbf{q}}\right)^{t}=\left(\begin{array}{cc}
\zeta_{\mathbf{q},A} & \zeta_{\mathbf{q},B}\end{array}\right)$, the mean-field Hamiltonian becomes block-diagonal in flavors as
follows

\begin{equation}
H_{\text{MF}}=\sum_{\mathbf{q}\in\frac{1}{2}\text{BZ}}\left(\begin{array}{cccccc}
\eta_{\mathbf{q}}^{x\dagger} & \eta_{\mathbf{q}}^{y\dagger} & \eta_{\mathbf{q}}^{z\dagger} & \theta_{\mathbf{q}}^{x\dagger} & \theta_{\mathbf{q}}^{y\dagger} & \theta_{\mathbf{q}}^{z\dagger}\end{array}\right)\left(\begin{array}{cccccc}
H_{\eta} & 0 & 0 & 0 & 0 & 0\\
0 & H_{\eta} & 0 & 0 & 0 & 0\\
0 & 0 & H_{\eta} & 0 & 0 & 0\\
0 & 0 & 0 & H_{\theta^{x}} & 0 & 0\\
0 & 0 & 0 & 0 & H_{\theta^{y}} & 0\\
0 & 0 & 0 & 0 & 0 & H_{\theta^{z}}
\end{array}\right)\left(\begin{array}{c}
\eta_{\mathbf{q}}^{x}\\
\eta_{\mathbf{q}}^{y}\\
\eta_{\mathbf{q}}^{z}\\
\theta_{\mathbf{q}}^{x}\\
\theta_{\mathbf{q}}^{y}\\
\theta_{\mathbf{q}}^{z}
\end{array}\right)+\text{constant,}\label{eq:HMF_matrix}
\end{equation}
in which
\begin{align}
H_{\theta^{\gamma}} & =2i\left(\begin{array}{cc}
0 & J_{\gamma}t_{i_{A}j_{B}}e^{-i\mathbf{q}\cdot\mathbf{n}_{\gamma}}\\
-J_{\gamma}t_{i_{A}j_{B}}e^{i\mathbf{q}\cdot\mathbf{n}_{\gamma}} & 0
\end{array}\right),\nonumber \\
H_{\eta} & =2i\sum_{\gamma}\left(\begin{array}{cc}
0 & J_{\gamma}u_{i_{A}j_{B}}^{\gamma}e^{-i\mathbf{q}\cdot\mathbf{n}_{\gamma}}\\
-J_{\gamma}u_{i_{A}j_{B}}^{\gamma}e^{i\mathbf{q}\cdot\mathbf{n}_{\gamma}} & 0
\end{array}\right).
\end{align}
The constant term will be henceforth neglected because it will not
affect the band structure and the mean-field dynamics. Two important
characteristics of the mean-field theory become apparent. First, there is no coupling among the flavors, which
allows us to write the ground state as the direct product 
\begin{equation}
\left|G\right\rangle =\left|G_{\eta^{x}}\right\rangle \otimes\left|G_{\eta^{y}}\right\rangle \otimes\left|G_{\eta^{z}}\right\rangle \otimes\left|G_{\theta^{x}}\right\rangle \otimes\left|G_{\theta^{y}}\right\rangle \otimes\left|G_{\theta^{z}}\right\rangle ,\label{eq:prod_GS}
\end{equation}
in which $\left|G_{\zeta}\right\rangle $ is the Fermi sea obtained
from the mean-field eigenstates of $H_{\zeta}$. Second, the $\theta^{\gamma}$
fermions will present flat bands with energy $\pm2\left|J_{\gamma}t_{i_{A}j_{B}}\right|$
and will be key to interpret the dynamical structure factor of the
model. 

\begin{figure}
\begin{centering}
\subfloat{\includegraphics[width=0.3\columnwidth]{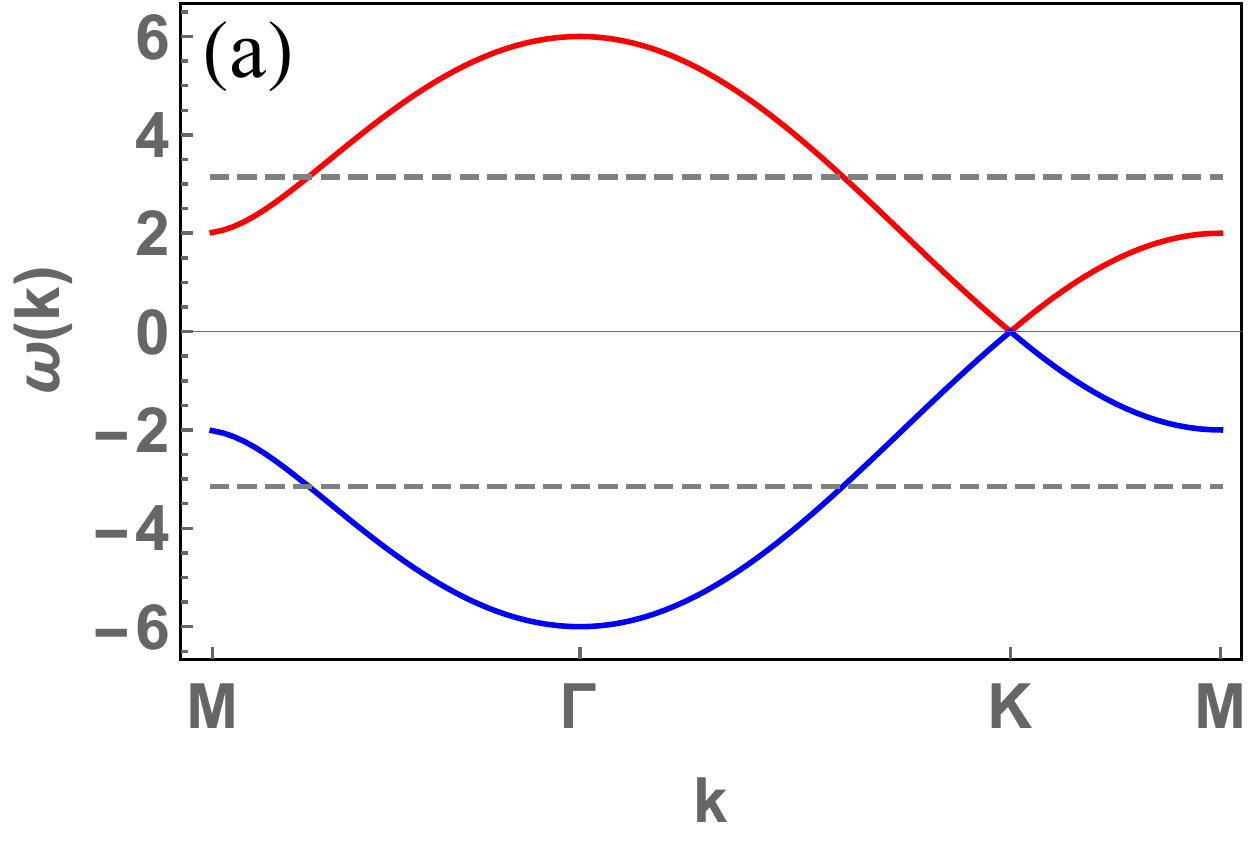}}\subfloat{\includegraphics[width=0.3\columnwidth]{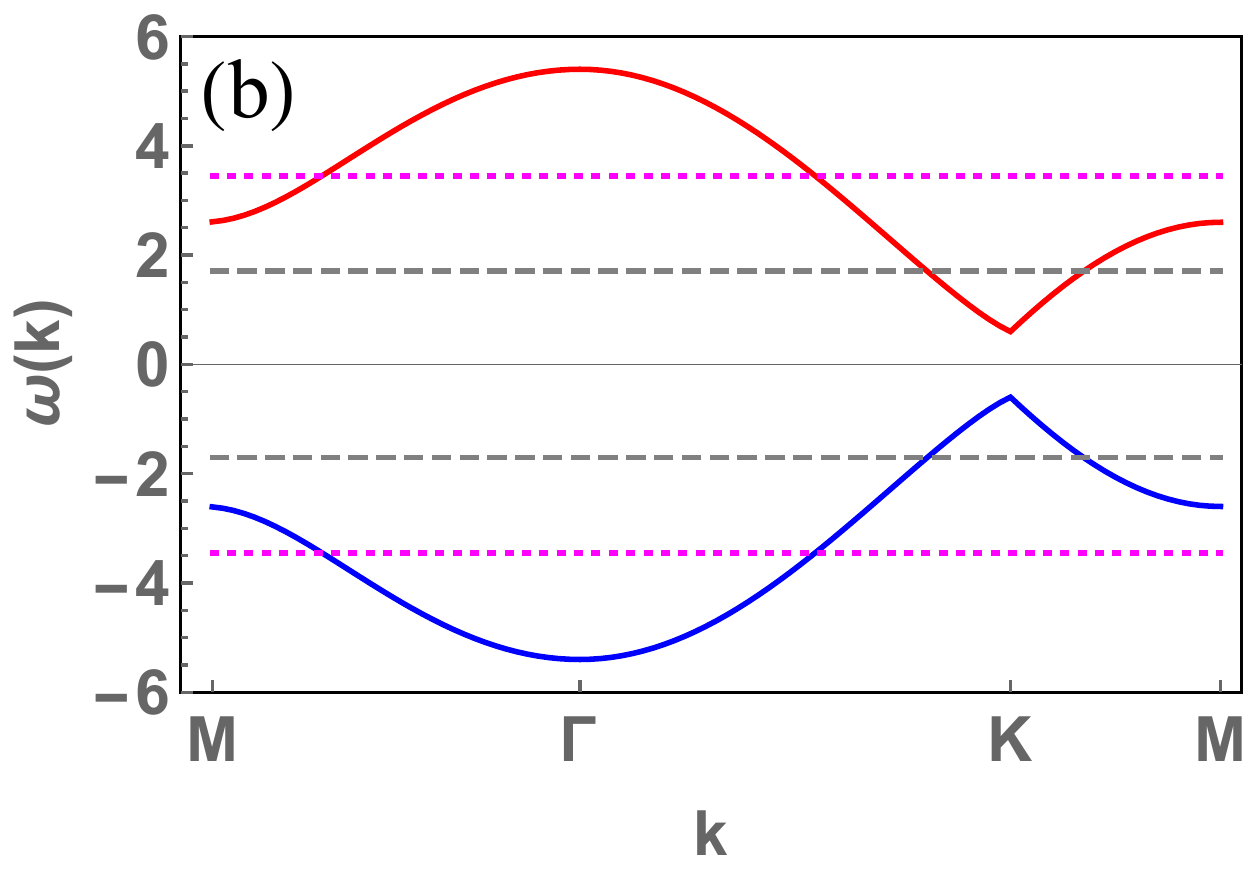}}\subfloat{\includegraphics[width=0.3\columnwidth]{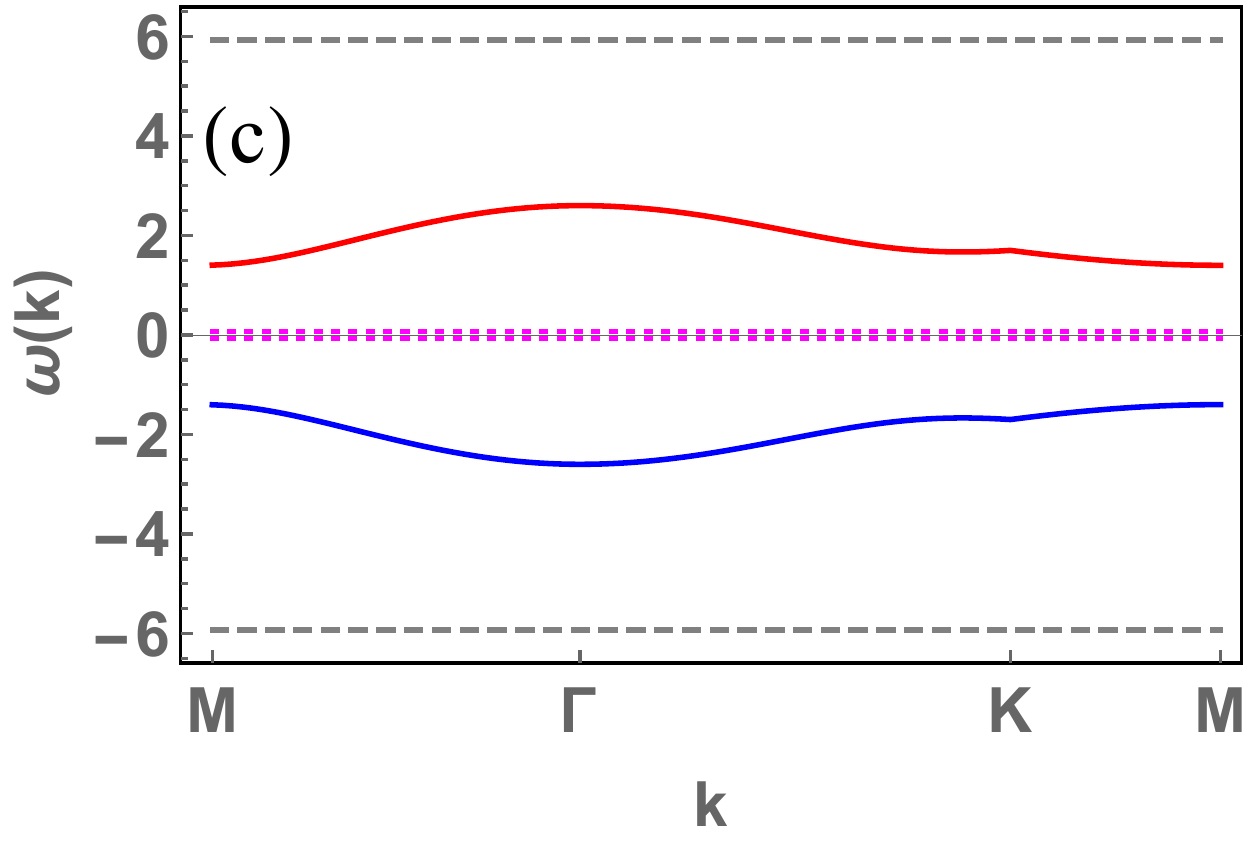}}
\par\end{centering}
\caption{\label{fig:First-Brillouin-zone} Mean-field band structure along
the high-symmetry lines of the Brillouin zone for three representative
cases: (a) the isotropic model $J_{x}=J_{y}=J_{z}$, (b) a gapless
model with $J_{x}=J_{y}$ and $J_{z}=0.7J_{x}$ and (c) a gapped case
with $J_{x}=J_{y}=0.15J_{z}$. The full lines represent the dispersion
of the $\eta$ fermions. Dashed lines represent the flat bands of
$\theta^{z}$ fermions, whereas dotted lines correspond to flat bands
for the $\theta^{x}$ or $\theta^{y}$. }
\end{figure}

We are now ready to find explicit expressions for the order parameters.
In general, we will be concerned with averages of the form $\left\langle i\zeta_{\mathbf{r},A}\zeta_{\mathbf{r}+\mathbf{n}_{\gamma},B}\right\rangle $,
which are explicitly given by
\begin{align}
\left\langle i\zeta_{\mathbf{r},A}\zeta_{\mathbf{r}+\mathbf{n}_{\gamma},B}\right\rangle  & =\frac{2i}{N}\sum_{\mathbf{q}\in\frac{1}{2}\text{BZ}}\left\langle e^{-i\mathbf{q}\cdot\mathbf{n}_{\gamma}}\zeta_{\mathbf{q},A}^{\dagger}\zeta_{\mathbf{q},B}-e^{i\mathbf{q}\cdot\mathbf{n}_{\gamma}}\zeta_{\mathbf{q},B}^{\dagger}\zeta_{\mathbf{q},A}\right\rangle \label{eq:general_MFT_parameter}
\end{align}
Without loss of generality, let us pick $t_{i_{A}j_{B}}>0$. The unitary
matrix that diagonalizes $H_{\theta^{\gamma}}$ is

\begin{align}
U_{\mathbf{q},\theta^{\gamma}} & =\frac{1}{\sqrt{2}}\left(\begin{array}{cc}
e^{i\mathbf{q}\cdot\mathbf{n}_{\gamma}/2} & ie^{-i\mathbf{q}\cdot\mathbf{n}_{\gamma}/2}\\
e^{i\mathbf{q}\cdot\mathbf{n}_{\gamma}/2} & -ie^{-i\mathbf{q}\cdot\mathbf{n}_{\gamma}/2}
\end{array}\right),
\end{align}
in which $U_{\theta^{\gamma}}H_{\theta^{\gamma}}U_{\theta^{\gamma}}^{\dagger}=2J_{\gamma}t_{i_{A}j_{B}}\text{diag}\left[\left\{ 1,-1\right\} \right]$.
The eigenstate operators are written like
\begin{equation}
\left(\begin{array}{c}
\Theta_{\mathbf{q},+}^{\gamma}\\
\Theta_{\mathbf{q},-}^{\gamma}
\end{array}\right)\equiv U_{\mathbf{q},\theta^{\gamma}}\left(\begin{array}{c}
\theta_{\mathbf{q},A}^{\gamma}\\
\theta_{\mathbf{q},B}^{\gamma}
\end{array}\right),\label{eq:Thetapm}
\end{equation}
and we find
\begin{equation}
\theta_{\mathbf{q}}^{\gamma\dagger}H_{\theta^{\gamma}}\theta_{\mathbf{q}}^{\gamma}=2J_{\gamma}t_{i_{A}j_{B}}\left(\Theta_{\mathbf{q},+}^{\gamma\dagger}\Theta_{\mathbf{q},+}^{\gamma}-\Theta_{\mathbf{q},-}^{\gamma\dagger}\Theta_{\mathbf{q},-}^{\gamma}\right).
\end{equation}
Following Eq. (\ref{eq:prod_GS}), we know that $u_{i_{A}j_{B}}^{\gamma}=\left\langle G_{\theta^{\gamma}}\left|i\theta_{\mathbf{r},A}^{\gamma}\theta_{\mathbf{r}+\mathbf{n}_{\gamma}}^{\gamma}\right|G_{\theta^{\gamma}}\right\rangle $.
The application of Eq. (\ref{eq:Thetapm}) on Eq. (\ref{eq:general_MFT_parameter})
then implies that
\begin{align}
u_{i_{A}j_{B}}^{\gamma} & =-1.
\end{align}
It is important to point out that this value of $u_{i_{A}j_{B}}^{\gamma}$
is independent of the value of the coupling constants $J_{\gamma}$
and leads to a ``hopping'' Hamiltonian of the $\eta$ fermions that
is the same as the zero-flux ground state of the exact solution. 

We are now ready to determine the order parameter $t_{i_{A}j_{B}}$.
For later reference, we define the functions $f_{x}(\mathbf{q})$,
$f_{y}(\mathbf{q})$ and $\phi(\mathbf{q})$
\begin{align}
f_{x}(\mathbf{q}) & =\sum_{\gamma}J_{\gamma}\cos\left(\mathbf{q}\cdot\mathbf{n}_{\gamma}\right),\nonumber \\
f_{y}(\mathbf{q}) & =\sum_{\gamma}J_{\gamma}\sin\left(\mathbf{q}\cdot\mathbf{n}_{\gamma}\right),\nonumber \\
\left|f(\mathbf{q})\right| & =\sqrt{f_{x}^{2}(\mathbf{q})+f_{y}^{2}(\mathbf{q})},\nonumber \\
e^{i\phi(\mathbf{q})} & =\frac{f_{x}(\mathbf{q})+if_{y}(\mathbf{q})}{\left|f(\mathbf{q})\right|}.
\end{align}
Notice that the spectrum of the three flavors of the $\eta$ fermions
is the same and given by $\epsilon_{\eta,\pm}(\mathbf{q})=\pm2\left|f(\mathbf{q})\right|$,
which recovers the dispersion of the fermionic magnons of the exact
solution. The unitary matrix that diagonalizes $H^{\eta}$ is 
\begin{align}
U_{\eta} & =\frac{1}{\sqrt{2}}\left(\begin{array}{cc}
e^{i\phi(\mathbf{q})/2} & -ie^{-i\phi(\mathbf{q})/2}\\
e^{i\phi(\mathbf{q})/2} & ie^{-i\phi(\mathbf{q})/2}
\end{array}\right),
\end{align}
in which the dependence with the coupling constants $J_{\gamma}$
is implicit in the phase factors $\phi(\mathbf{q})$. The parameter
$t_{i_{A}j_{B}}^{\gamma}$ is then given by
\begin{align}
t_{i_{A}j_{B}}^{\gamma} & =3\times\frac{2i}{N}\sum_{\mathbf{q}\in\frac{1}{2}\text{BZ}}\left\langle e^{-i\mathbf{q}\cdot\mathbf{n}_{\gamma}}\eta_{\mathbf{q},A}^{\dagger}\eta_{\mathbf{q},B}-e^{i\mathbf{q}\cdot\mathbf{n}_{\gamma}}\eta_{\mathbf{q},B}^{\dagger}\eta_{\mathbf{q},A}\right\rangle \nonumber \\
 & =\frac{6}{N}\sum_{\mathbf{q}\in\frac{1}{2}\text{BZ}}\cos\left[\phi\left(\mathbf{q}\right)-\mathbf{q}\cdot\mathbf{n}_{\gamma}\right].
\end{align}
The value of $t_{i_{A}j_{B}}^{\gamma}$ varies with the coupling constants
as indicated by the dependence with the phase $\phi\left(\mathbf{q}\right)$.
For the isotropic model, $t_{i_{A}j_{B}}^{\gamma}$ is the same for
all directions and a numerical calculation yields $t_{i_{A}j_{B}}\approx1.574$.
The evaluation of $t_{i_{A}j_{B}}^{\gamma}$ allows us to locate the
flat bands as indicated in Fig. \ref{fig:First-Brillouin-zone} for
the choices of $J_{\gamma}$ used as representative examples in the
main text.

\section{Dynamical Structure Factor at Mean-Field Level}

The main goal of this supplementary material is to provide the dynamical
structure factor of the SU(2)-symmetric model within mean-field theory.
We start from its expression in real space and time 

\begin{equation}
S_{lm}^{\alpha\alpha^{\prime}}(t)=\left\langle G\left|j_{l}^{\alpha}(t)j_{m}^{\alpha^{\prime}}(0)\right|G\right\rangle ,
\end{equation}
in which $j_{l}^{\alpha}$ is written like
\begin{equation}
j_{l}^{\alpha}=-\frac{1}{2}\sigma_{l}^{\alpha}-\sigma_{l}^{\alpha}T_{l}^{\left(\beta\gamma\right)}
\end{equation}
with $T_{l}^{\left(\beta\gamma\right)}$ given in Ref. \citep{Natori2017}.
The correlators of the form $\sigma_{l}^{\alpha}(t)\sigma_{m}^{\alpha}T_{m}^{\left(\beta\gamma\right)}(0)$
can be ignored in this mean field theory because such operators are
not flavor conserving and therefore vanish. Thus, $S_{lm}^{\alpha\alpha^{\prime}}(t)$
is given by the sum of the same matrix elements that were indicated
in the main text. In particular, the spin-spin correlations $I_{lm}^{\alpha\alpha^{\prime}}(t)=\left\langle G\left|\sigma_{l}^{\alpha}(t)\sigma_{m}^{\alpha^{\prime}}(0)\right|G\right\rangle $
within the Lehmann representation is given by
\begin{equation}
I_{lm}^{\alpha\alpha^{\prime}}(t)=-\sum_{\lambda}e^{i\left(E_{0}-E_{\lambda}\right)t}\left\langle G\left|\eta_{l}^{\beta}\eta_{l}^{\gamma}\right|\lambda\right\rangle \left\langle \lambda\left|\eta_{m}^{\beta^{\prime}}\eta_{m}^{\gamma^{\prime}}\right|G\right\rangle .
\end{equation}
Since we chose a mean-field decoupling that reproduces the exact ground
state and dispersion of the $\eta$ flavors, the function $I_{lm}^{\alpha\alpha^{\prime}}(t)$
computed within this approximation is the same as the exact solution.
An expression to the form factor of this dynamical response is discussed
in the main text, as well as the responses calculated for the three
representative cases.

Let us now discuss the dynamical correlation of spin-orbital operators

\begin{align}
W_{lm}^{\alpha\alpha^{\prime}}(t) & =\left\langle G\left|\sigma_{l}^{\alpha}T_{l}^{\left(\beta\gamma\right)}(t)\sigma_{m}^{\alpha^{\prime}}T_{m}^{\left(\beta^{\prime}\gamma^{\prime}\right)}(0)\right|G\right\rangle \nonumber \\
 & =-\sum_{\lambda}e^{i\left(E_{0}-E_{\lambda}\right)t}\left\langle G\left|\eta_{l}^{\alpha}\theta_{l}^{\left(\beta\gamma\right)}\right|\lambda\right\rangle \left\langle \lambda\left|\eta_{m}^{\alpha^{\prime}}\theta_{m}^{\left(\beta^{\prime}\gamma^{\prime}\right)}\right|G\right\rangle .
\end{align}
From the conservation of the number of flavors, we know that only
these three correlators should be evaluated 
\begin{align}
W_{lm}^{zz}(t) & =-\sum_{\lambda}e^{i\left(E_{0}-E_{\lambda}\right)t}\left\langle G\left|\eta_{l}^{z}\theta_{l}^{z}\right|\lambda\right\rangle \left\langle \lambda\left|\eta_{m}^{z}\theta_{m}^{z}\right|G\right\rangle \nonumber \\
W_{lm}^{xx}(t) & =-\frac{1}{4}\sum_{\lambda}e^{i\left(E_{0}-E_{\lambda}\right)t}\left\langle G\left|\eta_{l}^{x}\theta_{l}^{z}\right|\lambda\right\rangle \left\langle \lambda\left|\eta_{m}^{x}\theta_{m}^{z}\right|G\right\rangle \nonumber \\
 & -\frac{3}{4}\sum_{\lambda}e^{i\left(E_{0}-E_{\lambda}\right)t}\left\langle G\left|\eta_{l}^{x}\theta_{l}^{x}\right|\lambda\right\rangle \left\langle \lambda\left|\eta_{m}^{x}\theta_{m}^{x}\right|G\right\rangle \nonumber \\
W_{lm}^{yy}(t) & =-\frac{1}{4}\sum_{\lambda}e^{i\left(E_{0}-E_{\lambda}\right)t}\left\langle G\left|\eta_{l}^{y}\theta_{l}^{z}\right|\lambda\right\rangle \left\langle \lambda\left|\eta_{m}^{y}\theta_{m}^{z}\right|G\right\rangle \nonumber \\
 & -\frac{3}{4}\sum_{\lambda}e^{i\left(E_{0}-E_{\lambda}\right)t}\left\langle G\left|\eta_{l}^{y}\theta_{l}^{x}\right|\lambda\right\rangle \left\langle \lambda\left|\eta_{m}^{y}\theta_{m}^{x}\right|G\right\rangle .\label{eq:W_MFT}
\end{align}
One qualitative similarity between mean-field and exact solutions
of the Fourier transform of $W_{lm}^{zz}(t)$ is the presence of a
gapped response even when the spectrum of fermionic magnon is gapless.
However, the explanation for these gaps is very different. On the
exact level, the gap is caused by the formation of two gauge fluxes
and is determined by the vison gap $\Delta_{F}$ \citep{Knolle2014}.
On the mean-field level, the excitations of the $\theta^{\gamma}$
fermions will either create a hole on the negative energy flat band
or a particle on the positive energy one leading to an energy gap
of $2J_{\gamma}\left|t^{\gamma}\right|$.

\begin{figure}
\begin{centering}
\subfloat{\includegraphics[width=0.3\columnwidth]{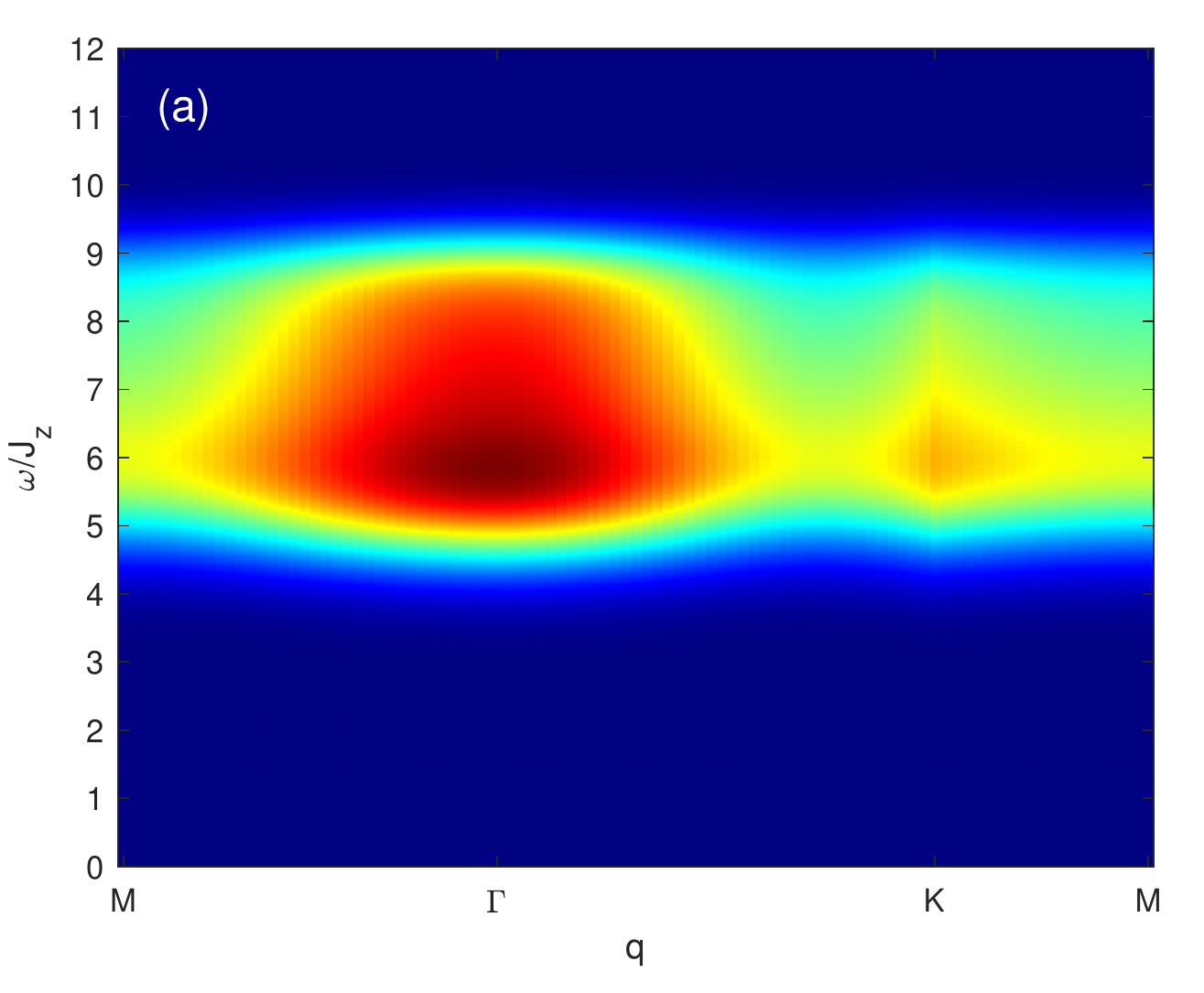}}\subfloat{\includegraphics[width=0.3\columnwidth]{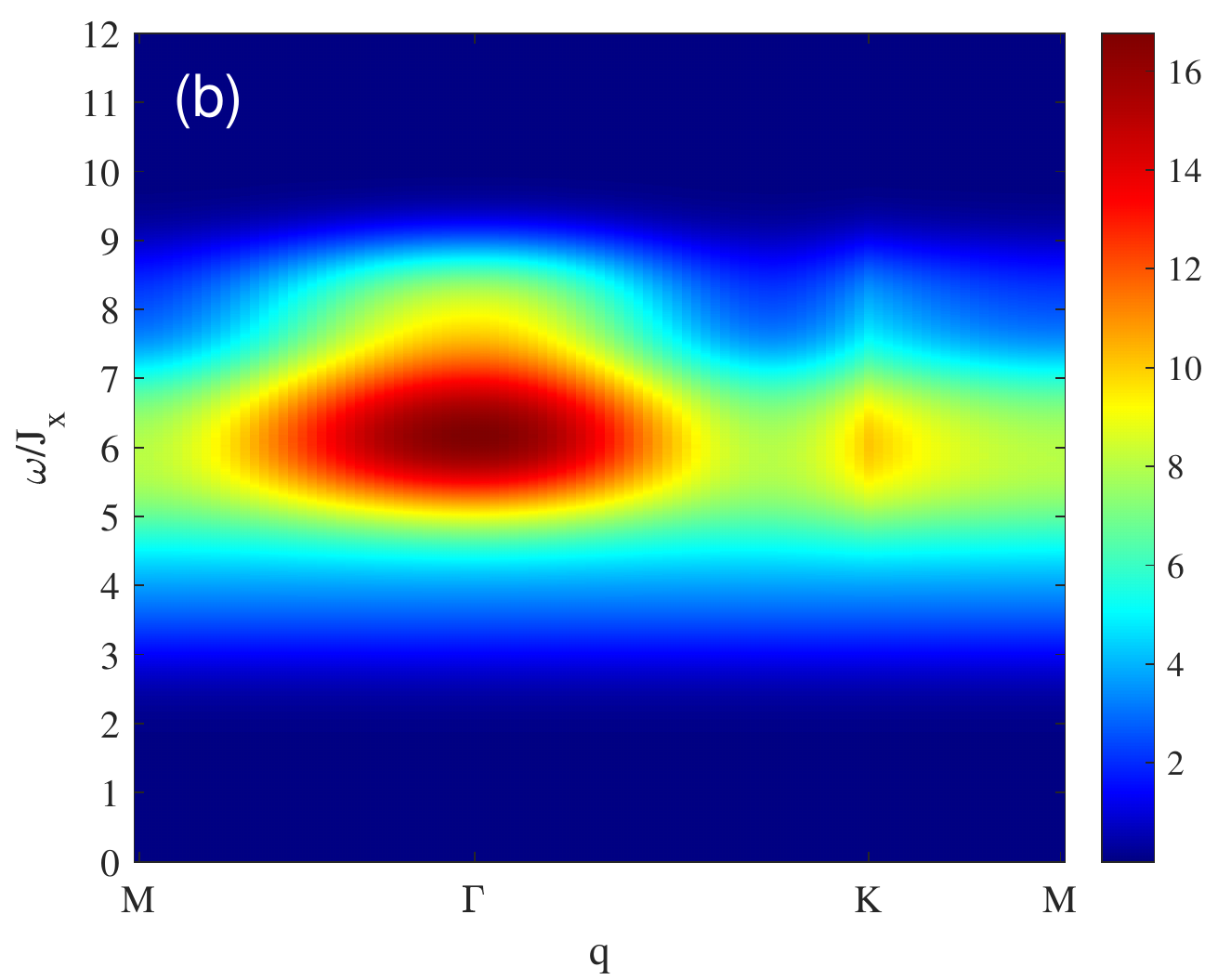}}\subfloat{\includegraphics[width=0.3\columnwidth]{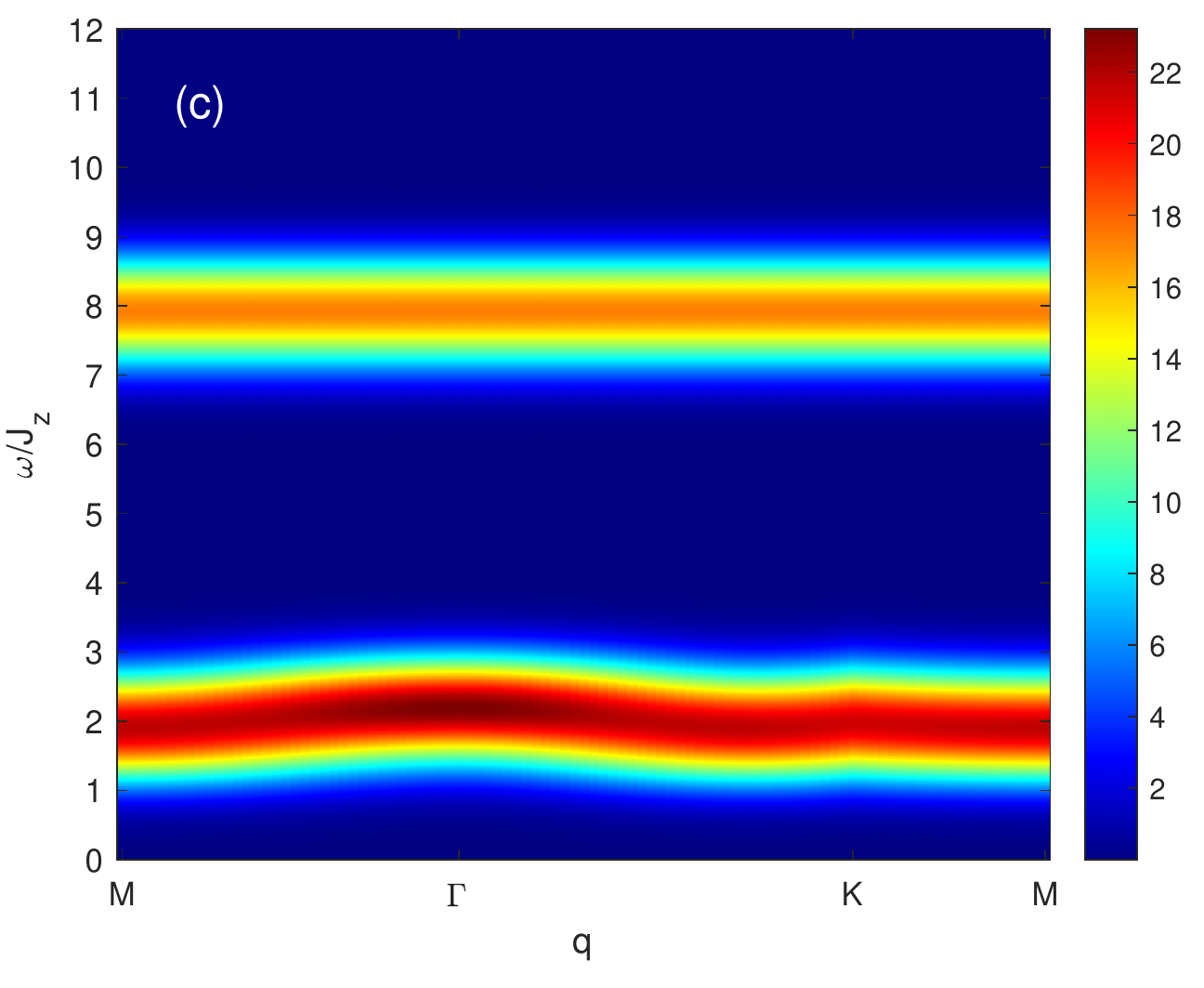}}
\par\end{centering}
\begin{centering}
\subfloat{\includegraphics[width=0.3\columnwidth]{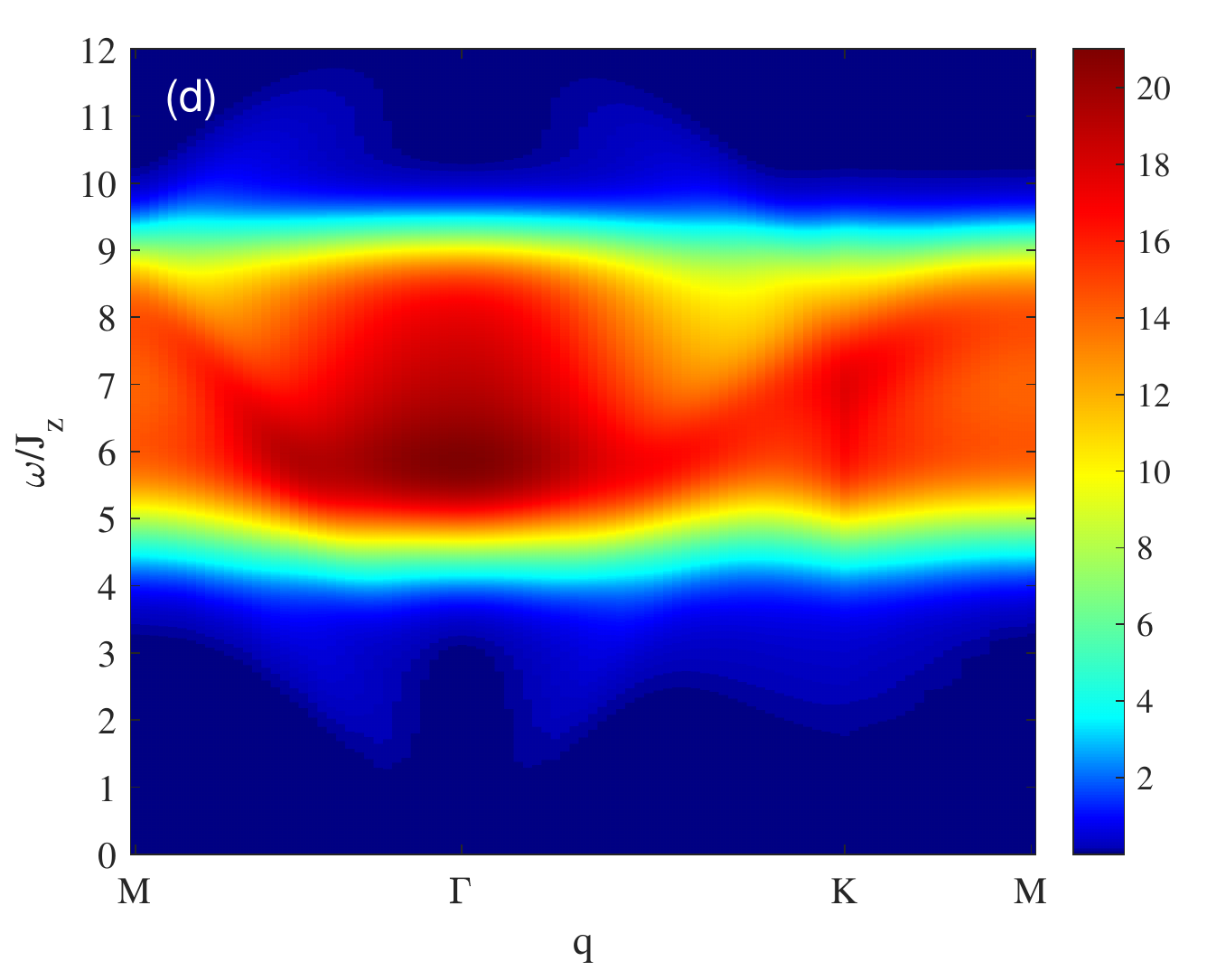}}\subfloat{\includegraphics[width=0.3\columnwidth]{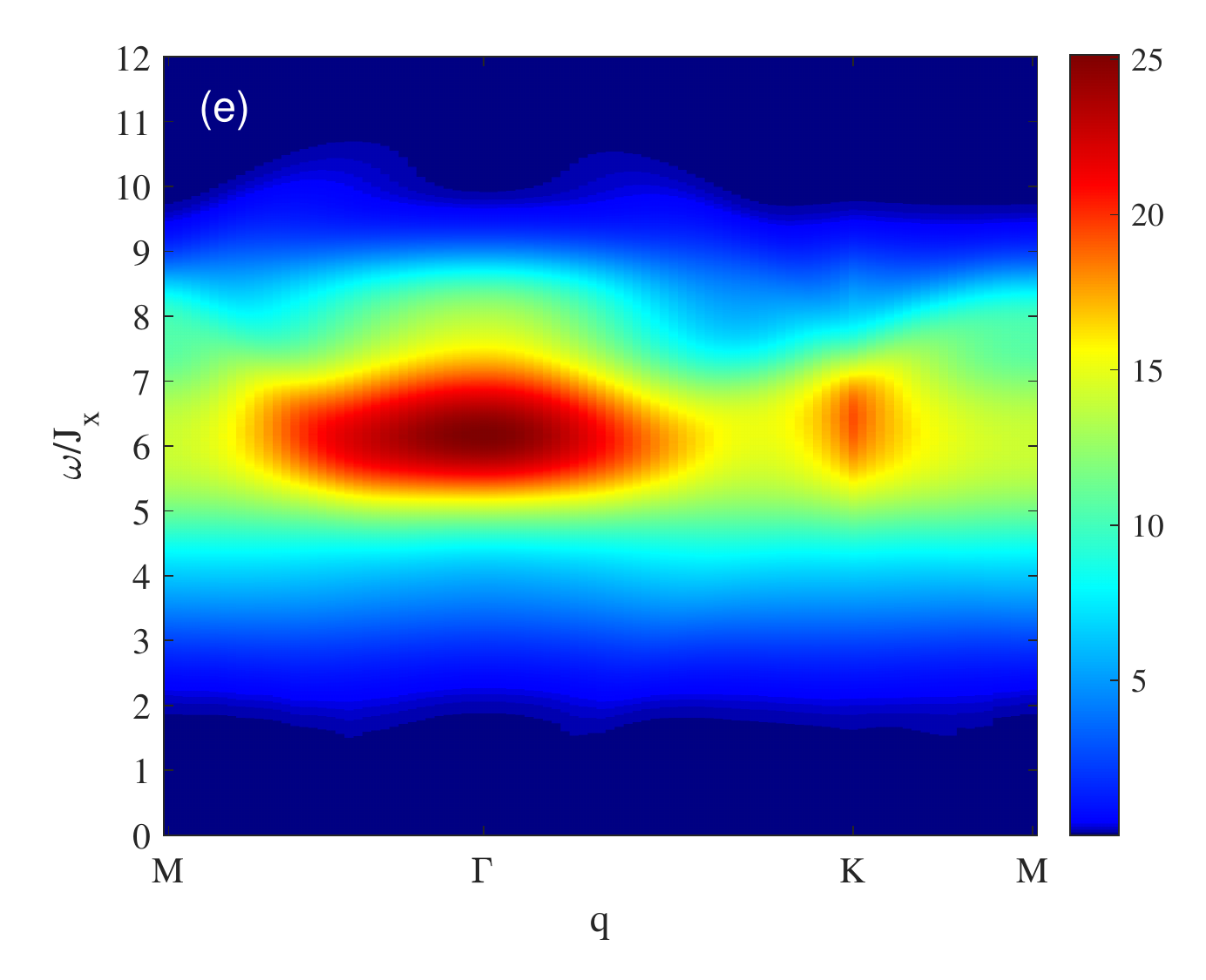}}\subfloat{\includegraphics[width=0.3\columnwidth]{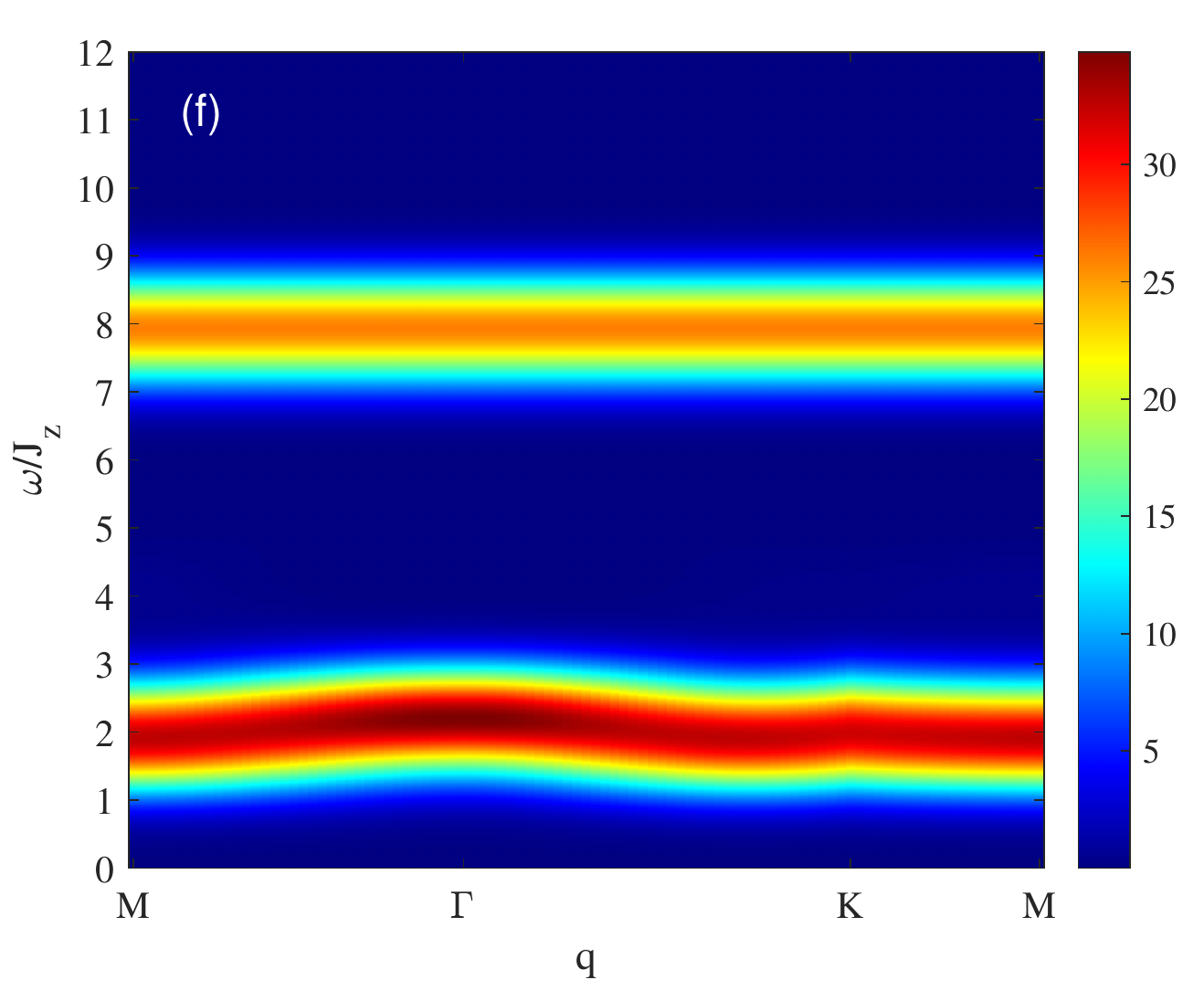}}
\par\end{centering}
\caption{\label{fig:MFT_DSF} The mean-field $W(\mathbf{q},\omega)$ and dynamical structure factor for (a-d) the isotropic
model with $J_{x}=J_{y}=J_{z}=1.0$, (b-e) a gapless but anisotropic
point $J_{x}=J_{y}$, $J_{z}=0.7J_{x}$ and (c-f) a gapped case with
$J_{x}=J_{y}=0.15J_{z}$.}
\end{figure}

The analytical formulas for the Fourier transform of Eq. (\ref{eq:W_MFT})
are

\begin{align}
W^{zz}(\mathbf{q},\omega) & =\frac{4\pi}{N}\sum_{\mathbf{q}_{1}\in\text{BZ}}\delta\left(\omega-2J_{z}\left|t_{z}\right|-2\left|f(\mathbf{q}_{1})\right|\right)\left[1+\cos\phi(\mathbf{q}_{1})\right],\\
W^{xx}(\mathbf{q},\omega) & =\frac{\pi}{N}\sum_{\mathbf{q}_{1}\in\text{BZ}}\delta\left(\omega-2J_{x}\left|t_{x}\right|-2\left|f(\mathbf{q}_{1})\right|\right)\left[1+\cos\phi(\mathbf{q}_{1})\right]\nonumber \\
 & +\frac{3\pi}{N}\sum_{\mathbf{q}_{1}\in\text{BZ}}\delta\left(\omega-2J_{x}\left|t_{x}\right|-2\left|f(\mathbf{q}_{1})\right|\right)\left[1+\cos\left[\phi(\mathbf{q}_{1})-\left(\mathbf{q}_{1}+\mathbf{q}\right)\cdot\mathbf{n}_{x}\right]\right]\\
W^{yy}(\mathbf{q},\omega) & =\frac{\pi}{N}\sum_{\mathbf{q}_{1}\in\text{BZ}}\delta\left(\omega-2J_{y}\left|t_{y}\right|-2\left|f(\mathbf{q}_{1})\right|\right)\left[1+\cos\phi(\mathbf{q}_{1})\right]\nonumber \\
 & +\frac{3\pi}{N}\sum_{\mathbf{q}_{1}\in\text{BZ}}\delta\left(\omega-2J_{y}\left|t_{y}\right|-2\left|f(\mathbf{q}_{1})\right|\right)\left[1+\cos\left[\phi(\mathbf{q}_{1})-\left(\mathbf{q}_{1}+\mathbf{q}\right)\cdot\mathbf{n}_{x}\right]\right].
\end{align}
The responses at mean-field level follow closely the density of states
translated by the flat-band gap $2J_{\gamma}\left|t_{\gamma}\right|$.
The cosine terms multiplying the density of states are related to
correlations between sites in two different sublattices, and account
for a small dependence of $W^{xx}$ and $W^{yy}$ on the transferred
momentum $\mathbf{q}$. We computed these expressions for the three
representative cases and presented their sum on Fig. \ref{fig:MFT_DSF},
which allows a comparison with the corresponding results presented
in the main text. In the exact response, the maximal spectral weight
occurs in the neighborhood of the flux gap $\Delta_{F}$, which is
generally incompatible with the flat-band gap $2J_{\gamma}\left|t^{\gamma}\right|$.
The maximum of the spectral weight within mean-field occurs near the
maximum of the density of states of the fermionic magnons shifted
by these flat-band gaps. As a result, the dynamics evaluated within
a mean-field approximation reproduces very poorly the exact results
and highlights the importance of the exact approach described in the main
text.

\end{document}